\documentclass[aip,pof,natbib]{revtex4-1}

\usepackage{graphicx}
\usepackage{dcolumn}
\usepackage{bm}

\usepackage[utf8]{inputenc}
\usepackage[T1]{fontenc}
\usepackage{mathptmx}
\usepackage{etoolbox}
\usepackage{xcolor}
\usepackage{url}
\usepackage[normalem]{ulem}
\usepackage{subfigure}
\usepackage{amsmath}
\usepackage{epstopdf, epsfig}

\makeatletter
\def\@email#1#2{%
 \endgroup
 \patchcmd{\titleblock@produce}
  {\frontmatter@RRAPformat}
  {\frontmatter@RRAPformat{\produce@RRAP{*#1\href{mailto:#2}{#2}}}\frontmatter@RRAPformat}
  {}{}
}%
\makeatother
\begin{document}


\title[Rotation-driven suppression and re-emergence of unsteady flow past a circular cylinder]{Rotation-driven suppression and re-emergence of unsteady flow past a circular cylinder
\\}

\author{Aditi Sengupta  \footnote{corresponding author}}
\email{aditi@iitism.ac.in}

\author{Sanjeev Kumar}

\author{Puneet Kumar}%

\affiliation{Department of Mechanical Engineering, Indian Institute of Technology Dhanbad, Jharkhand-826 004, India.
}%


\date{\today}

\begin{abstract}
The flow past a rotating cylinder undergoes substantial changes in its wake dynamics and aerodynamic response with increasing rotation rate. This study investigates these changes in two-dimensional compressible flow at \(Re_\infty=3800\) and \(M_\infty=0.1\), with the rotation rate systematically varied over \(0.1\leq\Omega\leq10\) using 130 high-fidelity simulations. At low rotation rates, conventional periodic vortex shedding is progressively suppressed with increasing rotation. A narrow transition regime around \(\Omega=2.2-2.8\) is identified, where sustained oscillations evolve into intermittent fluctuations before reaching a nearly quiescent state. At higher rotation rates, unsteadiness re-emerges with distinctly intermittent and multiscale characteristics, demonstrating that this instability is fundamentally different from conventional Kármán shedding. The surface pressure and streamwise skin-friction distributions reveal substantial redistribution of aerodynamic loading with rotation, characterized by increasingly intense suction and significant modification of the near-wall shear field. The maximum lift increases systematically with rotation, reflecting the strengthening Magnus-Robins effect, whereas drag exhibits a non-monotonic response with enhanced variability at high rotation rates. Despite the low free-stream Mach number, rotational acceleration produces local Mach numbers approaching 0.5, indicating appreciable compressibility effects in the near-cylinder region. Vorticity-based proper orthogonal decomposition further reveals a transition from wake-dominated coherent structures at low rotation to strongly wrapped, rotation-dominated structures at high rotation. These results establish a direct connection between rotational forcing, surface loading, compressibility, and the suppression and subsequent re-emergence of flow unsteadiness.
\end{abstract}

\maketitle

\section{\label{sec:level1}Introduction}

The flow past a rotating circular cylinder is a canonical problem for investigating the interaction between separated shear layers, rotation-induced vorticity and aerodynamic loading. In addition to its fundamental relevance to the Magnus–Robins effect, the problem has applications in flow control, rotating machinery, aerodynamics and propulsion. The flow is governed primarily by the Reynolds number and the ratio of cylinder surface speed to free-stream velocity, \(U_s/U_\infty\), with rotation producing a strong asymmetry in the boundary-layer development and wake structure. Consequently, steady rotation provides an effective means of modifying vortex shedding and the associated aerodynamic forces.

The influence of rotation rate on the incompressible cylinder wake has been investigated extensively using both experiments and numerical simulations. Early experiments by Tokumaru \& Dimotakis examined rotating cylinders over a relatively broad range of rotation rates and Reynolds numbers and demonstrated a strong departure of the measured mean lift from the classical Prandtl limit \citep{tokumaru1993lift}. Numerical studies subsequently established that rotation can progressively suppress the classical von Kármán shedding. Kang \textit{et al.} \citep{kang1999laminar}, for example, identified a critical rotation rate beyond which periodic shedding disappears at low Reynolds numbers. Mittal \& Kumar \citep{mittal2003flow} demonstrated a similar suppression at \(Re=200\), with vortex shedding disappearing at a rotation ratio of approximately \(U_s/U_\infty=1.9\). At still higher rotation rates, however, the wake does not evolve monotonically towards a steady state. Stojković \textit{et al.} \citep{stojkovic2002effect} identified a second shedding mode at high rotation rates, characterized by single-sided vortex shedding, while Kumar \textit{et al.} \citep{kumar2011flow} subsequently observed this mode experimentally. Global stability analyses by Pralits \textit{et al.} \citep{pralits2010instability,pralits2013three} further demonstrated that the classical von Kármán and single-sided shedding modes originate from distinct instability mechanisms. These studies establish that rotation acts not simply to attenuate wake unsteadiness, but can fundamentally reorganize the instability landscape of the cylinder wake.

The high-rotation regime is particularly relevant to the Magnus–Robins effect. Sengupta \textit{et al.} \citep{sengupta2003temporal} reported temporal instability at high rotation rates for \(Re=1000\) and 3800, demonstrating that a strongly rotating cylinder can develop unsteadiness even after conventional vortex shedding has been suppressed. Subsequent studies have examined the associated bifurcation and three-dimensional instability characteristics, while Rao \textit{et al.} \citep{rao2015review} summarized the range of two- and three-dimensional wake transitions observed with increasing rotation. Taken together, these investigations show that the response to rotation depends strongly on the operating regime. Conventional alternating shedding may be suppressed, followed by the appearance of asymmetric or single-sided structures and, at sufficiently high rotation, other forms of temporal instability.

Despite this extensive body of work, the majority of investigations have considered the incompressible formulation. This assumption becomes increasingly restrictive at high rotation rates because the velocity induced by the cylinder can greatly exceed the free-stream velocity. Consequently, the local Mach number near the cylinder may become appreciable even when the free-stream flow is nominally low-Mach-number. Compressibility can then modify the pressure field, boundary-layer development, vorticity dynamics and instability characteristics. The authors \cite{liu2023numerical, xue2024compressibility} reported compressibility-induced changes in shedding strength, Strouhal number and wake stability. However, compressibility does not always stabilize: at some parameter combinations compressibility can introduce or amplify other (acoustic or shear-layer) instabilities, especially as $M_{\infty}$ approaches transonic values or when shocklets/strong pressure gradients appear. Biringen {\it et al.} \cite{biringen1993numerical} noted that compressibility can stabilize or destabilize depending on rotational speeds, gap widths, and three-dimensional effects, for counter-rotating cylinders. Teymourtash \& Salimipour \citep{teymourtash2017compressibility} examined compressible flow over \(0.05\leq M_\infty\leq0.4\) and reported significant effects of compressibility on lift, drag and vortex shedding. Nagata {\it et al.} \cite{nagata2018direct} observed a similar trend in lift distribution for the compressible flow past a rotating sphere. Salimipour \& Anbarsooz \citep{salimipour2019surface} subsequently demonstrated that surface-temperature effects can further modify the compressible rotating-cylinder wake and its shedding mode. At substantially higher rotation rates, Sundaram \textit{et al.} \citep{sundaram2021multiscale} identified multiscale temporal instabilities in compressible flow at \(Re=3800\), while Suman \textit{et al.} \citep{suman2022novel} demonstrated that significant compressibility effects, including localized supersonic regions and weak shocks, can arise at \(M_\infty=0.1\) when the cylinder rotates sufficiently rapidly. These studies emphasize that free-stream Mach number alone is insufficient to characterize the compressibility of a highly rotating cylinder flow.

More recently, the present authors investigated the high-rotation regime at a fixed surface-speed ratio \(U_s/U_\infty=10\) and \(M_\infty=0.1\), while varying \(Re_\infty\) from 1000 to 6000 \citep{sengupta2025bifurcation}. The study identified the Reynolds-number-dependent onset of temporal instability and characterized the associated changes in the lift and drag dynamics. While this work established the sensitivity of the high-rotation Magnus–Robins instability to Reynolds number, it considered only a single, strongly rotating state. It therefore does not address how this high-rotation state emerges from the conventional cylinder wake as the rotation rate is increased. In particular, a systematic description connecting the low-rotation shedding regime, the intermediate suppression of unsteadiness and the subsequent high-rotation instability at a fixed Reynolds number remains lacking. 

The present study addresses this gap by examining two-dimensional compressible flow at \(Re_\infty=3800\) and \(M_\infty=0.1\) over a dense range of rotation rates, \(0.1\leq U_s/U_\infty\leq10\), comprising 130 high-fidelity simulations. Rather than focusing on a single instability regime, the objective is to establish how the flow field and aerodynamic response evolve continuously with rotation. Particular attention is given to the modification of the surface pressure and streamwise skin-friction distributions, the evolution of lift and drag, the emergence and suppression of temporal fluctuations, and the local compressibility generated by rotational acceleration. This systematic parameterization provides a direct connection between the classical wake at low rotation and the high-rotation Magnus–Robins instability identified in previous studies.

A further objective is to characterize how the coherent structures evolve across these flow regimes. Proper orthogonal decomposition (POD) has been extensively employed to identify dominant structures in cylinder wakes and to construct low-dimensional representations of their dynamics. For a canonical cylinder wake, POD modes typically organize into pairs associated with the phase evolution of large-scale vortex shedding, while higher modes represent progressively smaller-scale or higher-frequency structures. Enstrophy-based POD has also been used to relate the modal representation of the cylinder wake to instability dynamics and reduced-order modelling \citep{sengupta2015enstrophy}. The authors \cite{sengupta2016enstrophy} performed an enstrophy-based POD to the flow past a rotating cylinder at a supercritical rotation rate of 2.5 for a $Re_{\infty} = 60$ to explain the shedding of puffs of vortices. POD has also been applied to cylinders undergoing rotary oscillation. Ping \textit{et al.} \citep{ping2020wake}, for example, demonstrated that rotational forcing produces distinct wake responses whose dominant structures can be identified through POD. These studies establish the usefulness of POD for interpreting rotation-modified cylinder wakes, but have primarily considered low-to-moderate Reynolds numbers or oscillatory rotation.

To the best of our knowledge, a systematic POD characterization of a steadily rotating cylinder at \(Re_\infty=3800\) across a dense range of rotation rates extending from \(0.1\) to \(10\) has not been reported. Such an analysis is particularly valuable because the present parameter range spans fundamentally different dynamical states, from classical vortex shedding through suppression to intermittent high-rotation instability. In the present work, vorticity-based POD is therefore employed not merely as a data-reduction technique, but as a means of identifying the spatial organization underlying these transitions and relating the temporal activation of coherent structures to the observed aerodynamic dynamics.

The remainder of the paper is organized around this coupled description of flow regime, aerodynamic response and coherent-structure evolution. The results establish how rotation reorganizes the near-cylinder and wake flow and how this reorganization modifies surface loading and local compressibility. In doing so, the study provides a unified description of the rotation-induced flow dynamics of a compressible rotating cylinder over a substantially broader rotation-rate range than considered in previous high-fidelity compressible flow studies.
 
\section{Problem formulation of the rotating cylinder}

\begin{figure}[!ht]
\centering
\includegraphics[width=.9\textwidth]{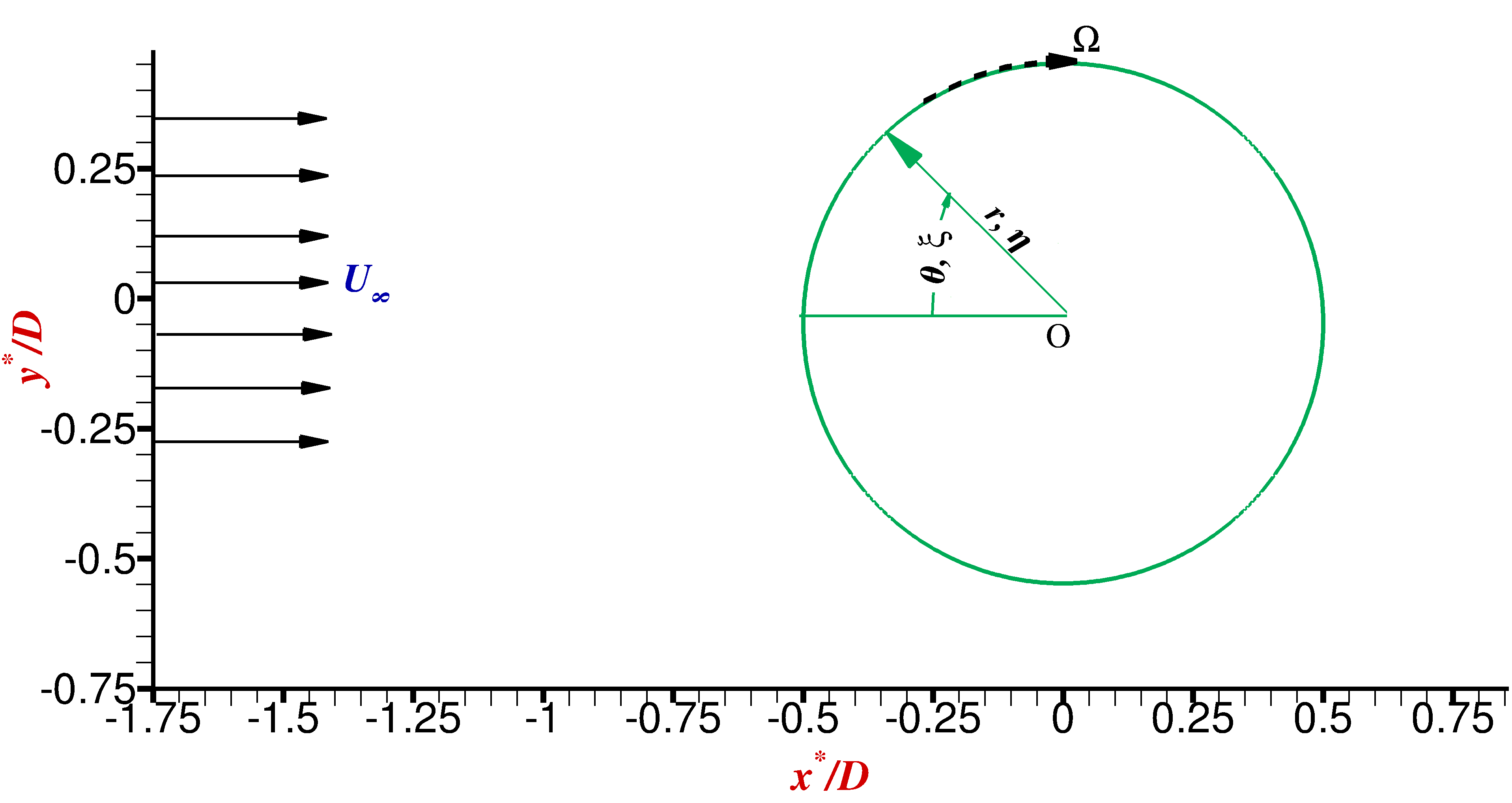}
\caption{Schematic of uniform flow past a rotating cylinder with surface speed, $U^*_s$.}
\label{fig1}
\end{figure}

The computational configuration considered in the present study is schematically illustrated in Fig. \ref{fig1}. A two-dimensional, uniform flow with free stream velocity, $U_\infty$ flows from left to right over an infinitely long circular cylinder of diameter, $D$. Owing to the assumed infinite span of the cylinder and the absence of spanwise variations, the problem is treated as two-dimensional. The cylinder rotates about its central axis with a prescribed dimensional angular velocity, $\Omega^*$, thereby imposing a tangential velocity on its surface. The Cartesian coordinate system $(x,y)$ is defined with its origin at the centre of the cylinder, with the $x$-axis aligned with the incoming free stream and the \(y\)-axis oriented normal to the free-stream direction. The polar coordinate $\theta$ is measured along the cylinder surface from the upstream stagnation point in the aziumthal direction, as indicated in Fig. \ref{fig1}. The imposed rotation generates a tangential surface velocity $U_s^*$, which is related to the dimensional angular velocity through $U_s^*=\Omega^*D/2$, or equivalently, $\Omega^*= 2U_s^*/D$. The strength of rotation is characterized using the nondimensional rotation rate, $ \Omega= U_s^*/U_\infty =\Omega^*D/2U_\infty$, which represents the ratio of the tangential velocity of the cylinder surface to the free-stream velocity. Thus, $\Omega=0$ corresponds to a stationary cylinder, whereas increasing values of $\Omega$ represent progressively stronger rotational forcing.

On the cylinder surface, the no-slip and impermeability conditions are imposed such that the fluid velocity matches the local tangential velocity of the rotating wall. The thermal boundary condition at the cylinder surface is taken to be adiabatic, implying zero heat flux in the direction normal to the wall, i.e., $\left.\frac{\partial T}{\partial n}\right|_{\mathrm{wall}}=0$, where $n$ denotes the outward normal direction to the cylinder surface. Consequently, the velocity and thermal boundary conditions are prescribed consistently with the rotating-wall configuration. The computational domain is extended sufficiently far from the cylinder to minimize the influence of the outer boundaries on the near-body flow and wake development. The far-field boundaries are located at a distance of 30$D$ from the cylinder, where characteristic-based boundary conditions \cite{pulliam1986implicit, joshi2023exploring} are employed. Such boundary treatment allows the propagation of incoming and outgoing acoustic and hydrodynamic disturbances while preventing spurious reflections from the computational boundaries, which is particularly important for the present compressible-flow simulations.

The incoming free stream is prescribed with uniform free stream conditions $(U_\infty,\rho_\infty,T_\infty)$, and the resulting flow is governed primarily by the Reynolds number and the nondimensional rotation rate. The Reynolds number, defined as $ Re=\frac{\rho_\infty U_\infty D}{\mu_\infty}$, is considered as 3800 while the inlet Mach number, $M_{\infty}$ is 0.1 (operating within the incompressibility limit) to replicate the experiments of \citet{tokumaru1993lift}. The present study systematically investigates the influence of the rotation parameter over the range, $ 0.1\leq\Omega\leq10$. This wide range of rotation rates enables the transition from weakly rotating wake dynamics to strongly rotation-dominated flow states to be examined systematically.

Having defined the physical configuration and the associated boundary conditions, the governing mathematical framework is introduced next. The flow field is described by the two-dimensional compressible Navier–Stokes equations, representing the conservation of mass, momentum, and total energy, as follows \cite{pal2026direct} 

\begin{eqnarray}
\frac{\partial {\hat{Q}}}{\partial t} + \frac{\partial {\hat{E}}}{\partial x} + \frac{\partial {\hat{F}}}{\partial y} = \frac{\partial {\hat{E_{v}}}}{\partial x} + \frac{\partial {\hat{F_{v}}}}{\partial y}
\label{eq1} 
\end{eqnarray}

\noindent where the conservative variables are given as, $\hat{Q} = [ \rho \;\; \rho u \;\; \rho v \;\; e_t]^T$. The convective flux vectors are similarly given as,

\begin{equation}
\hat{E} = [ \rho u \;\; \rho u^2 +p \;\; \rho uv \;\; (\rho e_t + p) u]^T 
\label{eq2} 
\end{equation}
\begin{equation}
\hat{F} = [ \rho v \;\; \rho uv \;\; \rho v^2 +p \;\; (\rho e_t + p) v]^T
\label{eq3} 
\end{equation}

\noindent and the viscous flux vectors are given as

\begin{equation}
\hat{E_{v}}= [ 0\;\; \tau _{xx} \;\; \tau _{xy} \;\; (u \tau _{xx} +v \tau _{xy} -q _{x})]^T
\label{eq4} 
\end{equation}
\begin{equation}
\hat{F_{v}}= [0 \;\; \tau _{yx} \;\; \tau _{yy} \;\; (u \tau _{yx} +v \tau _{yy} -q _{y})]^T
\label{eq5} 
\end{equation}

In Eqs. \eqref{eq1}–\eqref{eq5}, \(\rho\), \(u\), \(v\), \(e_t\), and \(p\) represent the nondimensional density, streamwise and transverse velocity components, total specific energy, and pressure, respectively. The governing variables are nondimensionalized using the free-stream quantities \(\rho_\infty\), \(U_\infty\), \(T_\infty\), and \(\mu_\infty\) as the reference density, velocity, temperature, and dynamic viscosity, respectively. The cylinder diameter \(D\) is used as the characteristic length scale, while \(D/U_\infty\) defines the characteristic time scale. The resulting formulation is governed by dimensionless parameters of free-stream Reynolds number (\(Re_\infty\)), Mach number (\(M_\infty\)), and Prandtl number (\(Pr\)), which are defined as 
\vspace{-1mm}
\begin{equation*}
Pr = \frac{\mu_{\infty} C_{p}}{\kappa}	;\ Re_{\infty} = \frac{\rho_{\infty} U_{\infty} D}{\mu_{\infty}}	;\ M_{\infty} = \frac{U_{\infty}}{\sqrt{\gamma R^* T_{\infty}}}
\end{equation*}    

\noindent Here, \(\gamma=1.4\) denotes the ratio of the specific heats at constant pressure and constant volume, i.e., \(\gamma=C_p/C_v\). The reference pressure is evaluated from the ideal-gas relation as \(p_\infty=\rho_\infty R^*T_\infty\), where \(R^*\) is the dimensional specific gas constant. Accordingly, its nondimensional form is defined as \(R=R^*T_\infty/U_\infty^2\). The thermodynamic state variables are related through the ideal-gas equation of state, $p=\rho RT$, while the total specific energy is expressed as $e_t=C_vT+\frac{1}{2}\left(u^2+v^2\right)$. The heat-flux components in the streamwise and transverse directions are evaluated using Fourier's law of heat conduction as
\vspace{-2mm}
\begin{equation*}
q_{x}= - \frac{\mu}{Pr Re_{\infty} (\gamma -1) M_{\infty}^2} \frac{\partial {T}}{\partial x}; \; \;\\
q_{y}= - \frac{\mu}{Pr Re_{\infty} (\gamma -1) M_{\infty}^2} \frac{\partial {T}}{\partial y}
\end{equation*}

\noindent The components of the Newtonian viscous stress tensors are expressed as
\vspace{-1mm}
\begin{equation*}
\tau_{xx} = \frac{1}{Re_{\infty}}\Big[ 2\mu \frac{\partial u}{\partial x} + \lambda \nabla \cdot \vec{V} \Big]; \; \\
\tau_{yy} = \frac{1}{Re_{\infty}}\Big[2\mu\frac{\partial v}{\partial y}+ \lambda \nabla \cdot \vec{V} \Big]; \; \\
\tau_{xy} = \tau_{yx} = \frac{\mu}{Re_{\infty}}\Big[\frac{\partial u}{\partial y} + \frac{\partial v}{\partial x} \Big]
\end{equation*}

\noindent Here, \(\tau_{xx}\), \(\tau_{yy}\), \(\tau_{xy}\), and \(\tau_{yx}\) denote the components of the viscous stress tensor. The bulk viscosity coefficient \(\lambda\) is evaluated using Stokes' hypothesis, namely, $\lambda=-\frac{2}{3}\mu$, while the dynamic viscosity \(\mu\) is evaluated as a function of temperature using Sutherland's law. To accurately represent the curved geometry of the rotating cylinder, the governing equations are subsequently transformed from the physical Cartesian coordinate system \((x,y)\) to a body-fitted computational coordinate system \((\xi,\eta)\), defined through the coordinate transformations, $ \xi=\xi(x,y), \eta=\eta(x,y)$. The resulting governing equations in the transformed computational plane are then expressed in strong conservation form as

\begin{eqnarray}
\frac{\partial {{\bf Q}}}{\partial t} + \frac{\partial {\bf E}}{\partial \xi} + \frac{\partial {\bf F}}{\partial \eta} = \frac{\partial {\bf E_{v}}}{\partial \xi} + \frac{\partial {\bf F_{v}}}{\partial \eta}
\label{eq6} 
\end{eqnarray}   

\noindent with the state variables and flux vectors, given as 
\begin{equation*}
{\bf Q} = \hat{Q}/J; \; \; \; \\
{\bf E} = (\xi _{x} \hat{E} +\xi _{y} \hat{F})/J; \; \; \; \\
{\bf F} = (\eta _{x} \hat{E} +\eta _{y} \hat{F})/J; \; \; \;\\
{\bf E_{v}} = (\xi _{x} \hat{E_{v}} +\xi _{y} \hat{F_{v}})/J; \; \; \;\\
{\bf F_{v}} = (\eta _{x} \hat{E_{v}} +\eta _{y} \hat{F_{v}})/J
\end{equation*}

\noindent and $J$ is the Jacobian of the grid transformation given by $
J=\frac{1}{x_{\xi} y_{\eta} -x_{\eta} y_{\xi}}$. The grid metrics \(\xi_x\), \(\xi_y\), \(\eta_x\), and \(\eta_y\) are evaluated with generation of the body-fitted \(O\)-type grid using the hyperbolic grid-generation procedure in Pointwise. These metrics are defined as $\xi_x=Jy_\eta, \; \xi_y=-Jx_\eta, \; \eta_x=-Jy_\xi,\;\eta_y=Jx_\xi$, and are subsequently employed to transform the spatial derivatives from the physical Cartesian plane \((x,y)\) to the computational plane \((\xi,\eta)\). For the computational domain shown in Fig. \ref{fig1}, an \(O\)-type grid comprising 401 uniformly distributed points in the circumferential (\(\xi\)) direction and 450 points in the radial (\(\eta\)) direction is employed. To adequately resolve the near-wall flow region, the grid points in the radial direction are clustered towards the cylinder surface using the following hyperbolic tangent stretching function:

$$ r(\eta)=0.5+r_{\max}\left[1- \frac{\tanh\left\{\beta\left(\dfrac{N_\eta-\eta}{N_\eta-1}\right)\right\}} {\tanh(\beta)}\right], $$

\noindent where \(N_\eta\) denotes the total number of grid points in the \(\eta\)-direction, \(r_{\max}\) corresponds to the location of the far-field boundary, which is set to \(30D\) in the present simulations, and \(\beta\) is the stretching parameter controlling the degree of grid clustering near the cylinder surface and, consequently, the wall-normal grid spacing \(\Delta r_{\mathrm{wall}}\). A grid-independence study for the rotating-cylinder configuration was previously reported by \citet{sundaram2021multiscale}, in which three near-wall resolutions, namely \(\Delta r_{\mathrm{wall}}=0.0005D\), \(0.001D\), and \(0.002D\), were considered. The lift coefficient, \(C_l\), obtained using the two finest near-wall resolutions was found to be essentially identical. Based on this assessment, the present simulations employ the finest wall-normal resolution, \(\Delta r_{\mathrm{wall}}=0.0005D\), corresponding to a stretching parameter of \(\beta=3\).

\subsection{Numerical methodology}
The governing equations in Eq. \eqref{eq6} are solved using a high-order finite-difference framework. The spatial derivatives of the convective fluxes are discretized using the optimized upwind compact scheme, OUCS3, which provides near-spectral resolution over a broad range of wavenumbers \citep{sagaut2023global}. The derivatives associated with the viscous fluxes are evaluated using a non-uniform explicit central-difference scheme. To suppress unresolved high-wavenumber numerical oscillations, a fourth-order artificial diffusion term following \citet{jameson2017origins} is incorporated with a diffusion coefficient of \(0.015\). Temporal integration is performed using the optimized three-stage Runge–Kutta scheme (OCRK3) \citep{sagaut2023global}, with a nondimensional time step of $\Delta t=1.25\times10^{-5}$. The coupled spatial and temporal discretization has previously been examined using global spectral analysis, demonstrating its ability to preserve the dispersion characteristics associated with the relevant spatial and temporal scales of the flow. The simulations are performed using an in-house finite-difference compressible NSE solver parallelized using the message passing interface. Domain decomposition is implemented using the non-overlapping high-accuracy parallel (NOHAP) algorithm \citep{sundaram2023non, sundaram2022non}, which eliminates numerical errors arising at subdomain interfaces without requiring an extensive overlap between adjacent subdomains. The accuracy of the parallel implementation has been extensively assessed for a range of incompressible and compressible flow configurations \cite{joshi2024highly, joshi2025comparing, sengupta2024separation, sengupta2024thermal}, demonstrating agreement with the corresponding sequential computations to machine precision.

The present study investigates the influence of cylinder rotation on the dynamics and stability characteristics of compressible flow past a circular cylinder. The nondimensional rotation rate is systematically varied over the range $0.1 \leq \Omega \leq 10$, while the free-stream Reynolds number is maintained at $Re_{\infty}=3800$. The free-stream Mach number is fixed at $M_{\infty}=0.1$. The Prandtl number is prescribed as \(Pr=0.71\), representative of air and reference free-stream temperature and density are taken as \(T_\infty=288.15~\mathrm{K}\) and \(\rho_\infty=1.2256~\mathrm{kg/m^3}\), respectively. A total of 130 high-fidelity simulations are performed to systematically resolve the evolution of the flow over a wide range of rotation rates. The rotation rate is sampled at intervals of \(0.1\) over \(0.1\leq\Omega\leq7.0\), while a finer increment of \(0.05\) is employed over \(7.0<\Omega\leq10\). This sampling strategy provides a high-resolution numerical database for examining the progressive modification of the wake dynamics as the relative importance of rotational forcing increases. At higher rotation rates, the drag coefficient, \(C_d\), exhibits increasingly pronounced variations with relatively small changes in the rotation rate. Consequently, a finer sampling resolution is employed for \(\Omega>7\) to adequately capture the rapid variation in the aerodynamic response.

The numerical methodology employed in the present work has previously been assessed against the experimental measurements of \citet{tokumaru1993lift} for flow past a rotating cylinder \citep{sundaram2021multiscale, sengupta2025bifurcation}. \citet{sundaram2021multiscale} performed simulations at \(Re_\infty=3800\) for surface rotation speeds ranging from \(0.5U_\infty\) to \(10U_\infty\), and the computed mean lift coefficients showed reasonable agreement with the corresponding experimental measurements. Sengupta {\it et al.} \cite{sengupta2025bifurcation} further showed that the simulated transverse velocity component for a rotation rate of $\Omega =10$ and $Re_{\infty} = 3800$ matched the corresponding experimental measurement of \citet{tokumaru1993lift}. The present simulations employ the same validated numerical framework to extend the analysis towards a systematic investigation of rotation-induced modifications to the flow field and its associated dynamical behaviour.

Each simulation is advanced sufficiently long to capture the transient evolution and subsequent long-time flow behaviour. The resulting database is subsequently used to examine the dependence of the aerodynamic response and flow dynamics on the rotation rate. The complete set of 130 simulations constitutes a high-fidelity numerical database for compressible flow past a rotating cylinder at fixed $Re_{\infty}$ and $M_{\infty}$, with a total computational cost of approximately \(1.3\) million core-hours. This extensive dataset provides a detailed characterization of the transition from weakly rotating wake dynamics to strongly rotation-dominated flow states.

\section{Results and discussion}
In this section we provide a systematic examination of the influence of cylinder rotation on the flow characteristics over the range \(0.1\leq \Omega \leq10\). The discussion begins with the evolution of the instantaneous flow structures and the corresponding surface pressure and skin-friction distributions. The associated aerodynamic response is subsequently analysed through the temporal evolution and spectral characteristics of the lift coefficient, together with the variation of key global quantities, including the maximum lift and drag coefficients and the onset time of flow unsteadiness. Finally, the underlying flow dynamics and the evolution of coherent structures are investigated through vorticity-based modal decomposition, providing a comprehensive characterization of the rotation-induced changes in the wake dynamics.

\subsection{Dynamics of compressible flow past a weakly and strongly rotating cylinder}

\begin{figure*}[!ht]
\centering
\includegraphics[width=\textwidth]{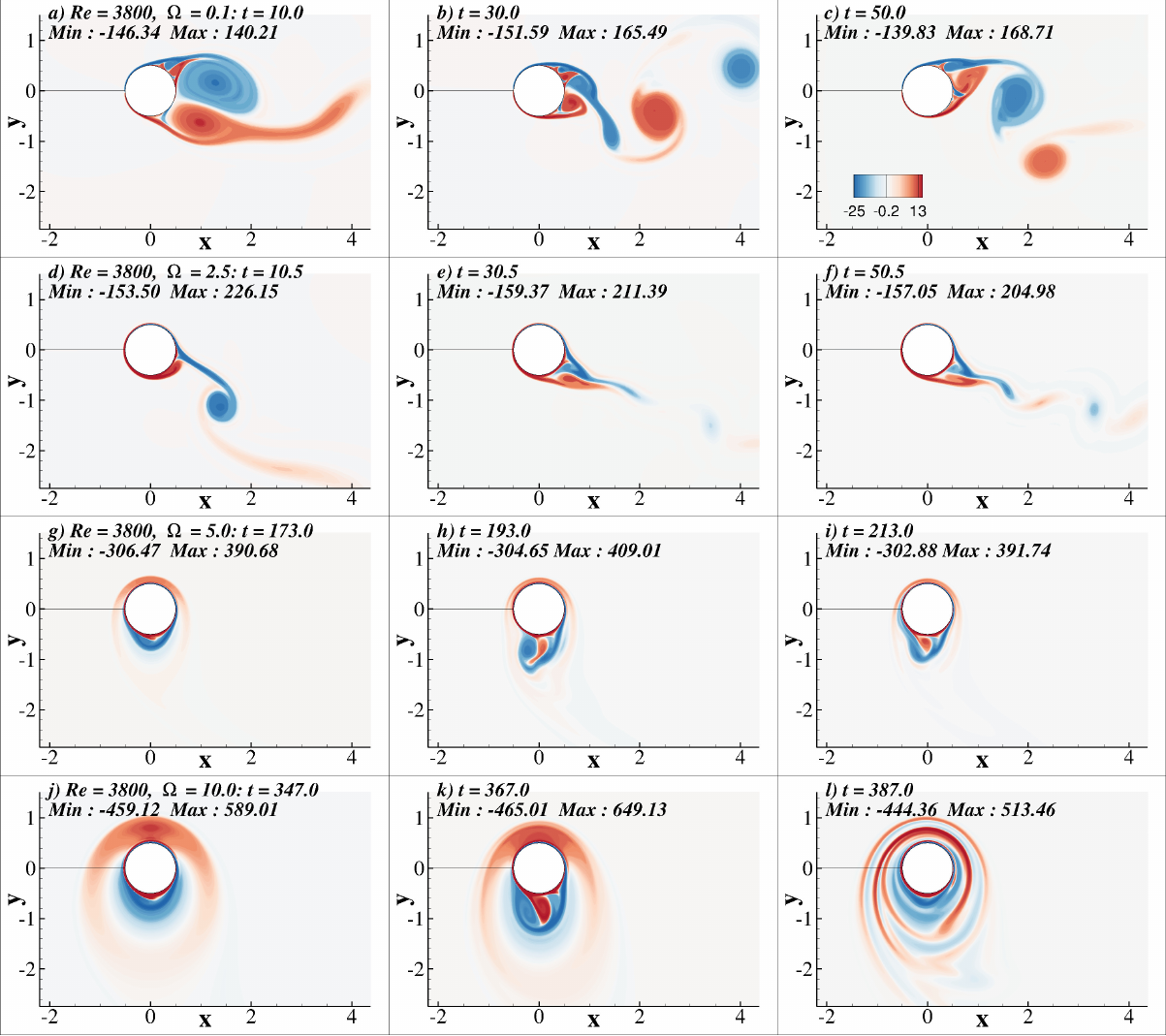}
\caption{Contours of vorticity shown for rotation rates, $\Omega = 0.1$, 2.5, 5.0 and 10.0 in frames (a)-(c), (d)-(f), (g)-(i), and (j)-(l), respectively. The time-instants are chosen such that leftmost frames correspond to onset time of the instability.}
\label{fig2}
\end{figure*}

Figure \ref{fig2} presents instantaneous vorticity contours at three successive time instants after the onset of instability for four representative rotation rates, \(\Omega=0.1\), \(2.5\), \(5.0\), and \(10.0\), at \(Re_\infty=3800\) and \(M_\infty=0.1\). The selected cases illustrate the progressive modification of wake dynamics with increasing rotational forcing, spanning the conventional rotating-cylinder wake regime, suppression of large-scale vortex shedding, and the emergence of temporally evolving high-rotation flow structures associated with the Magnus–Robins effect \cite{kumar2011flow, rao2015review}. 

For the lowest rotation rate shown in Figs. \ref{fig2}(a–c), the flow exhibits the formation and downstream convection of vortical structures of alternating sign. The wake retains characteristics of the conventional vortex-shedding process, although the imposed rotation introduces an asymmetry between the two sides of the cylinder and modifies the development of the separated shear layers \cite{sengupta2023compressibility, sengupta2020effects, sengupta2020effectsa}. Such asymmetric modification of the classical cylinder wake by rotation has been extensively documented. Numerical simulations by \citet{mittal2003flow} demonstrated that the progressive increase in rotation alters the vortex-shedding process and modifies the topology of the near wake, while experimental investigations by \citet{lam2009vortex} reported corresponding changes in the vortex-shedding characteristics of rotating cylinders at $Re_{\infty}$ comparable to the present configuration. Thus, the vortical evolution observed at \(\Omega=0.1\) remains consistent with the weak-rotation regime in which the underlying wake instability is still closely related to the classical von Kármán-type shedding mechanism.

A substantial modification of the wake is observed at \(\Omega=2.5\), depicted in Figs. \ref{fig2}(d–f). Compared with the \(\Omega=0.1\) case, the vortical structures are considerably weaker and the large-scale wake activity is strongly reduced. The vorticity generated at the cylinder surface remains confined predominantly to the immediate near-wake region, with only weak structures subsequently transported downstream. This progressive attenuation and eventual suppression of conventional vortex shedding with increasing rotation is consistent with the literature \cite{mittal2003flow, kumar2011flow} wherein a critical rotation range was identified beyond which the classical alternating vortex street is suppressed. \citet{lam2009vortex} reported progressive weakening of vortex shedding over \(3600\le Re\le5000\), with strong suppression at rotation rates approaching the present transition range. Although the precise critical rotation rate depends on $Re_{\infty}$ and flow configuration, the present \(\Omega=2.5\) case clearly lies within a regime in which the conventional wake dynamics are strongly altered by rotational forcing.

The flow undergoes a further qualitative reorganization at \(\Omega=5\) shown in Figs. \ref{fig2}(g–i). Rather than forming a conventional alternating vortex street, the vorticity remains strongly concentrated in the vicinity of the cylinder and undergoes pronounced temporal evolution around the rotating body. The surface-generated vorticity is transported by the strong tangential motion of the cylinder, leading to compact vortical structures that remain closely coupled to the near-body flow. This behaviour represents a departure from the classical wake-shedding mechanism and is more appropriately interpreted in the context of the high-rotation dynamics of the Magnus–Robins effect \cite{sengupta2003temporal}. The high-rotation vortical dynamics observed here are consistent with multiscale instability mechanisms \cite{sundaram2021multiscale, suman2022novel} for compressible flow past a rotating cylinder at high rotation rates.

At \(\Omega=10\), depicted in Figs. \ref{fig2}(j–l), the influence of rotation becomes dominant, and the instantaneous vorticity field exhibits a markedly different vortical distribution from that observed at lower rotation rates. The vorticity remains concentrated around the cylinder and progressively develops into extended curved layers surrounding the body. At later times, the oppositely signed vorticity structures assume a spiral-like arrangement around the cylinder, indicating that rotational transport and near-body vorticity dynamics dominate over the conventional downstream vortex-shedding process. The emergence of such temporally evolving high-rotation structures is consistent with the broader class of Magnus–Robins instabilities \cite{sengupta2003temporal} which were subsequently investigated for compressible flow \cite{kumar2026supervised, sengupta2025bifurcation}. 

\begin{figure}[!ht]
\centering
\includegraphics[width=.8\textwidth]{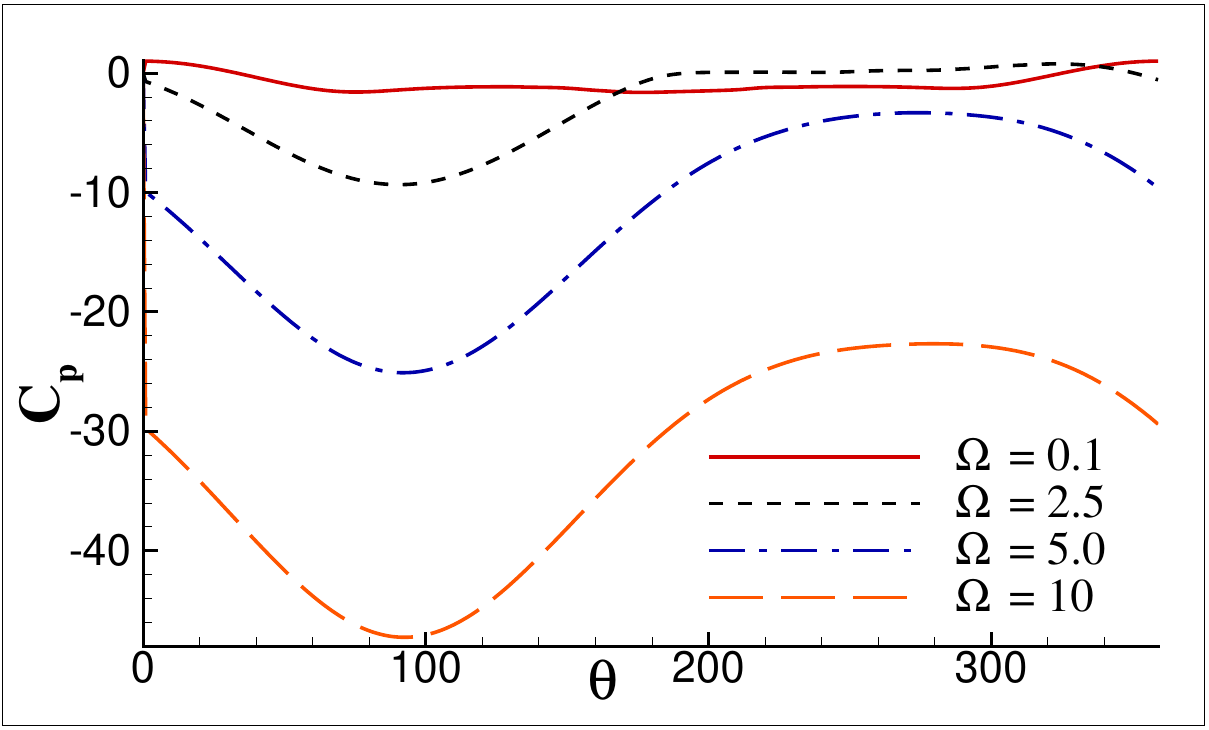}
\caption{Variation of coefficient of pressure, $C_{p}$ along the surface of the cylinder, prescribed by $\theta$, for rotation rates, $\Omega = 0.1$, 2.5, 5.0 and 10.0.}
\label{fig3}
\end{figure}

Figure \ref{fig3} shows the time-averaged pressure coefficient, \(C_p\), distributed along the cylinder surface as a function of the azimuthal coordinate \(\theta\) for rotation rates, \(\Omega=0.1\), \(2.5\), \(5.0\), and \(10.0\). A pronounced and systematic modification of the mean surface pressure field is observed with increasing rotation rate. In particular, the magnitude of the negative $C_p$ increases substantially as \(\Omega\) is increased, indicating the progressively stronger influence of rotational motion on the near-surface flow and the associated pressure loading. The distributions shown in Fig. \ref{fig3} provide a direct representation of the mean pressure mechanism underlying the rotation-induced aerodynamic forces.

For the lowest rotation rate, \(\Omega=0.1\), the time-averaged $C_p$ remains relatively weak, varying only modestly around the cylinder circumference. At this rotation rate, the tangential velocity imposed by the cylinder surface is small compared with the free-stream velocity, and the mean pressure field consequently retains the characteristics of a weakly perturbed cylinder flow. At low rotation rates, the flow is primarily governed by the conventional separated wake dynamics \cite{sengupta2019direct, sundaram2020effects, sengupta2020nonmodal, sengupta2018enstrophy}, with rotation introducing only a moderate asymmetry between the two sides of the cylinder.

A substantial change occurs at \(\Omega=2.5\), for which a pronounced region of mean suction develops, with the minimum $C_p$ reaching approximately \(-10\). The location of this minimum corresponds to the portion of the surface where the imposed rotational motion strongly modifies the velocity of the external flow. Rotation accelerates the fluid within the near-wall region on the rotation-augmented side, producing a sustained reduction in the mean static pressure. Conversely, the pressure recovers over the remaining portion of the circumference, thereby generating a substantial azimuthal pressure asymmetry \cite{ohkura1997pressure}. The resulting pressure difference between the two sides of the cylinder constitutes the principal contribution to the Magnus force.

The effect becomes considerably more pronounced for \(\Omega=5\) and \(10\). The minimum value of $C_p$ decreases to approximately \(-25\) and \(-47\), respectively, demonstrating that the mean surface pressure field becomes increasingly dominated by the imposed rotational velocity. It is important to note that \(C_p\) is normalized using the free-stream dynamic pressure. Consequently, as the tangential surface velocity increases according to $U_s=\Omega U_\infty$, the characteristic velocity scales associated with the rotation-induced flow can become several times larger than \(U_\infty\). The resulting pressure differences can therefore greatly exceed the pressure scale associated with the translational free stream alone, leading to values of $C_p$ whose magnitude is substantially greater than unity. The extremely large negative values observed for \(\Omega=5\) and \(10\) should thus be interpreted as a consequence of the strong rotation-induced acceleration of the surrounding flow, rather than as an anomalous pressure response.

At sufficiently large \(\Omega\), the rotational velocity becomes the dominant mechanism governing the near-body momentum distribution. The mean pressure field consequently undergoes a fundamental reorganization, with a large portion of the aerodynamic loading arising from the strongly asymmetric pressure distribution established around the rotating cylinder. This behaviour provides a near-body explanation for the large Magnus forces reported in experiments and simulations of rapidly rotating cylinders \cite{miller1976technique, sundaram2021multiscale}.

\begin{figure}[!ht]
\centering
\includegraphics[width=.8\textwidth]{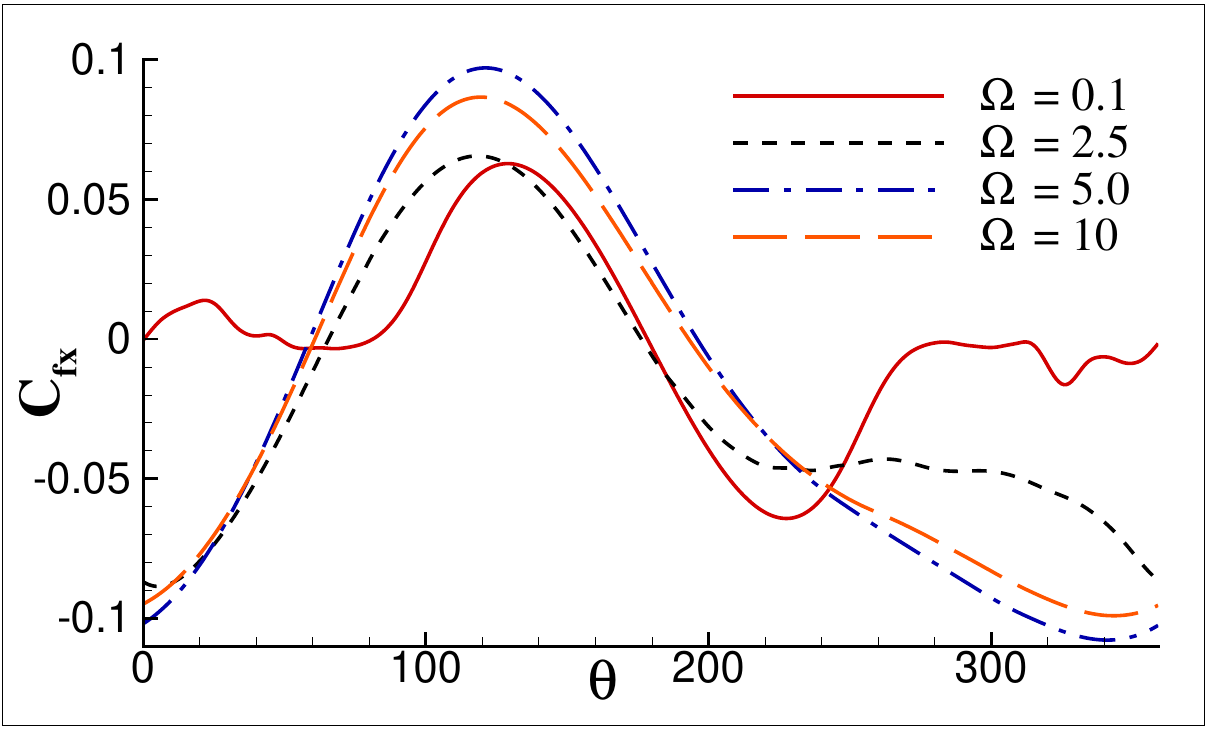}
\caption{Variation of coefficient of skin friction, $C_{f}$ along the surface of the cylinder, prescribed by $\theta$, for rotation rates, $\Omega = 0.1$, 2.5, 5.0 and 10.0.}
\label{fig4}
\end{figure}

Figure \ref{fig4} presents the time-averaged skin-friction coefficient, \(C_{f}\), as a function of the azimuthal coordinate \(\theta\) for rotation rates, \(\Omega=0.1\), \(2.5\), \(5.0\), and \(10.0\). The distributions demonstrate that cylinder rotation substantially modifies the mean viscous loading around the circumference. As the rotation rate increases, the skin-friction distribution develops progressively larger positive and negative extrema, indicating an increasing contribution of the imposed tangential wall motion to the near-wall momentum balance \cite{sengupta2021role, sengupta2016roles, sengupta2023role}.

At the lowest rotation rate, \(\Omega=0.1\), the distribution is strongly influenced by the translational free stream and the associated development of the boundary layer around the cylinder. The skin friction remains relatively small over much of the circumference and exhibits several changes in sign. These sign changes are associated with the local evolution of the near-wall velocity gradient which in turn reflect changes in the projection of the surface shear vector onto the streamwise direction. The comparatively weak magnitude of $C_{f}$ is consistent with a flow in which the rotational velocity represents only a small perturbation to the conventional boundary-layer and separated-wake dynamics.

A pronounced modification occurs as the rotation rate is increased to \(\Omega=2.5\). The magnitude of the positive skin-friction peak increases substantially, with the maximum occurring over the portion of the cylinder where the local rotational motion strongly influences the near-wall flow. At the same time, an extended region of negative \(C_{f}\) develops over the downstream part of the cylinder. The resulting distribution reflects the strong circumferential transport of momentum induced by the rotating wall \cite{glauert1957boundary}. Unlike a stationary cylinder, for which the wall shear is generated solely by the external flow and boundary-layer development, the rotating cylinder continuously injects tangential momentum into the fluid. The relative importance of this rotational forcing increases with \(\Omega\), producing progressively stronger wall-normal velocity gradients and hence larger viscous stresses.

The same trend becomes more evident for \(\Omega=5\) and \(10\). The positive peak of \(C_{f}\) increases further, while the negative values over the latter portion of the circumference become increasingly pronounced. At these rotation rates, the near-wall flow is no longer governed primarily by the translational free stream; instead, the tangential motion of the cylinder becomes a dominant component of the local velocity field. The imposed rotation continuously drags the adjacent fluid along the cylinder surface, generating intense near-wall shear \cite{sengupta2023multi, sengupta2023effects, sengupta2022thermally} and substantially modifying the circumferential distribution of vorticity. 

Another important feature of Fig. \ref{fig4} is the systematic displacement of the locations of the extrema and zero crossings with increasing rotation rate. These changes indicate that rotation alters not only the magnitude of the wall shear but also the angular distribution of the near-wall momentum exchange. Experimental evidence for this behaviour was reported \cite{aldoss1990experimental}, wherein the authors directly measured $C_{f}$, stagnation-point location, and separation characteristics for a rotating cylinder in crossflow. It was reported that skin friction is strongly dominated by moving-wall effect on the upstream-moving side of the cylinder, while rotation substantially shifts the stagnation point and produces an asymmetric wake.

\begin{figure}[!ht]
\centering
\includegraphics[width=\textwidth]{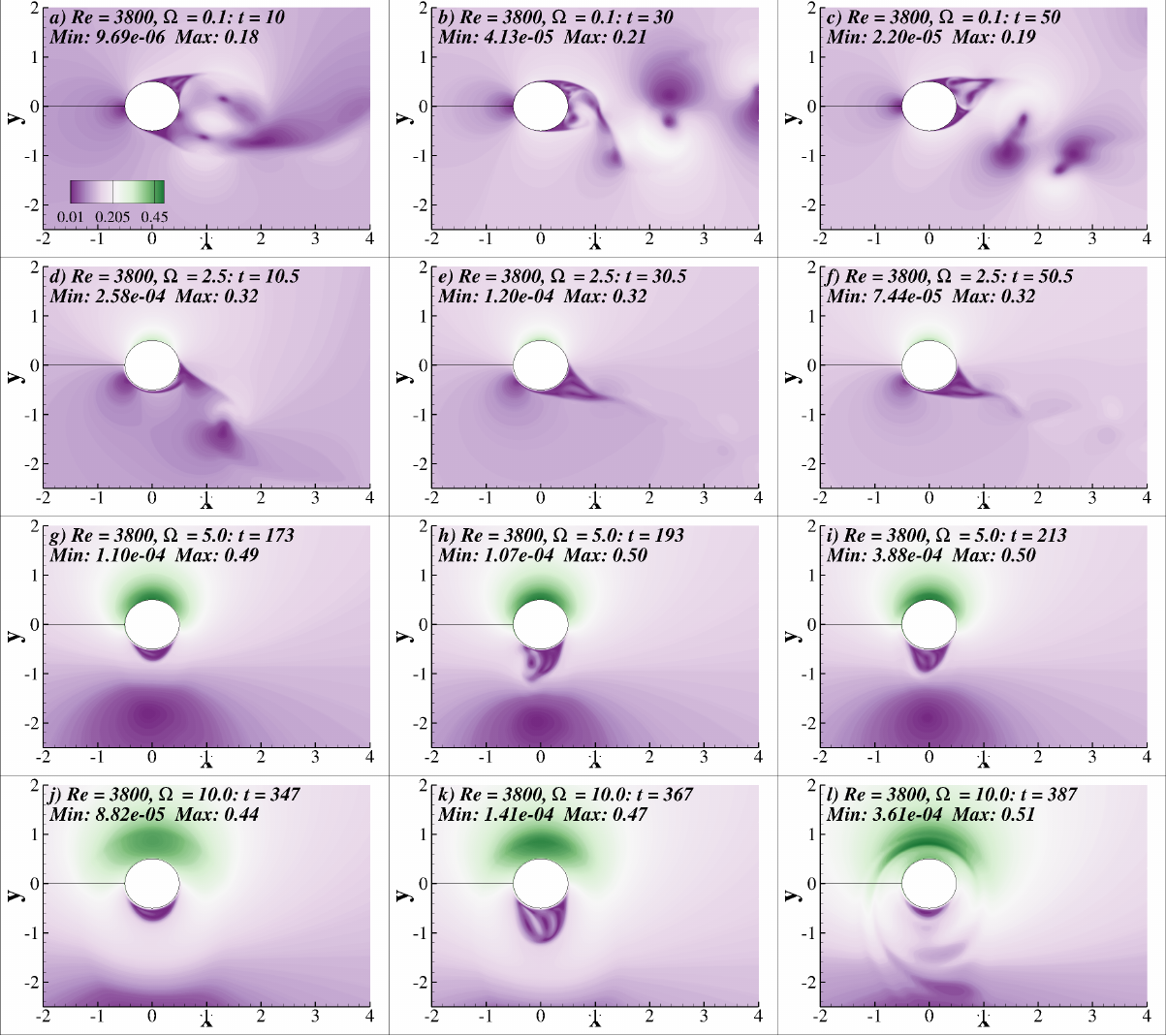}
\caption{Contours of Mach number shown for rotation rates, $\Omega = 0.1$, 2.5, 5.0 and 10.0 in frames (a)-(c), (d)-(f), (g)-(i), and (j)-(l), respectively. The time-instants are chosen such that leftmost frames correspond to onset time of the instability.}
\label{fig5}
\end{figure}

Figure \ref{fig5} presents the instantaneous Mach-number contours at three successive time instants, beginning from the onset of instability, for rotation rates, \(\Omega=0.1\), \(2.5\), \(5.0\), and \(10.0\), at \(Re_\infty=3800\) and \(M_\infty=0.1\). While the corresponding vorticity fields in Fig. \ref{fig2} have already established the rotation-induced evolution of the wake, the present figure examines how these flow structures modify the local compressibility of the flow. 

For the lowest rotation rate, \(\Omega=0.1\) shown in Figs. \ref{fig5}(a–c), the local Mach number remains relatively modest, with the maximum value varying between approximately \(0.18\) and \(0.21\) over the three time instants. The regions of elevated Mach number are primarily associated with the acceleration of the flow around the cylinder and the subsequent development of the unsteady wake. Since the rotational surface velocity is small compared with the free-stream velocity, the local velocity field is governed predominantly by the translational motion of the incoming flow. Consequently, although the wake undergoes temporal evolution, the departure of the local Mach number from \(M_\infty=0.1\) remains limited.

A considerably stronger local amplification of Mach number is observed for \(\Omega=2.5\) depicted in Figs. \ref{fig5}(d–f). The maximum Mach number increases to approximately \(0.32\), indicating that rotation-induced acceleration has become an important component of the local flow dynamics. The regions of elevated Mach number are concentrated around the cylinder and within the asymmetric shear layer extending into the wake. The imposed tangential surface motion modifies the local velocity through direct momentum transfer to the fluid adjacent to the cylinder. As this rotationally accelerated fluid interacts with the free stream, strong spatial gradients develop in the near-body region and along the separated shear layers. Thus, even though the incoming flow remains weakly compressible, the local velocity scale is no longer determined solely by \(U_\infty\).

The effect becomes more pronounced for \(\Omega=5\). The maximum local Mach number reaches approximately \(0.49-0.50\), as shown in Figs. \ref{fig5}(g–i), corresponding to nearly a fivefold increase relative to $M_{\infty}$. The high-Mach-number regions are concentrated predominantly around the upper portion of the cylinder, where the rotational motion produces strong acceleration of the surrounding fluid. Simultaneously, the lower region of the cylinder and the near wake exhibit distinct low-Mach-number zones associated with the asymmetric redistribution of momentum and the evolving vortical structures. The resulting Mach-number field therefore closely reflects the strong asymmetry of the rotation-dominated flow identified in the vorticity contours of Fig. \ref{fig2}.

At the highest rotation rate, \(\Omega=10\) shown in Figs. \ref{fig5}(j–l), the maximum Mach number varies from approximately \(0.44\) at the onset of instability to values approaching \(0.51\) during the subsequent evolution. The regions of locally elevated Mach number become more spatially extensive and remain closely associated with the curved, rotation-dominated flow surrounding the cylinder. Importantly, the maximum Mach number does not increase linearly with \(\Omega\), despite the continuous increase in the prescribed surface speed. This suggests that the local Mach number is determined by the nonlinear interaction between the imposed rotation, viscous diffusion, pressure redistribution, and the evolving instability, rather than by the rotational velocity alone.

The physical origin of the enhanced local compressibility can be understood by considering the characteristic velocity scales of the problem. The free-stream Mach number is defined using only the incoming velocity, $M_\infty=\frac{U_\infty}{a_\infty}=0.1$, whereas the rotating cylinder introduces an additional tangential velocity scale, $U_s=\Omega U_\infty$. Thus, local flow velocity in the vicinity of the cylinder can differ substantially from \(U_\infty\). The rotational wall motion continuously transfers momentum to the surrounding fluid and generates regions of strong acceleration, particularly where the rotational and externally induced motions reinforce each other. Since the Mach number is a local quantity, an increase in local velocity, together with local variations in temperature and hence the speed of sound, can produce appreciable departures from $M_{\infty}$.

Earlier incompressible simulations \cite{mittal2003flow, pralits2010instability} demonstrated that rotation can substantially reorganize the wake and produce a second instability regime at sufficiently high rotation rates. However, such an incompressible formulation cannot capture the accompanying density–pressure coupling associated with strong local acceleration. The importance of explicitly accounting for compressibility in rapidly rotating-cylinder flows was subsequently emphasized \cite{teymourtash2017compressibility, salimipour2019surface, nagata2018direct}. It is noteworthy that the present maximum Mach numbers remain below unity for all four rotation rates. Therefore, the flow remains globally subsonic, and the contours do not indicate the formation of shock waves. Nevertheless, increase from \(M_\infty=0.1\) to local values approaching \(M\simeq0.5\) is physically significant. 

\subsection{Aerodynamic loading and scale selection for low and high rotation rates}

\begin{figure}[!ht]
\centering
\includegraphics[width=.85\textwidth]{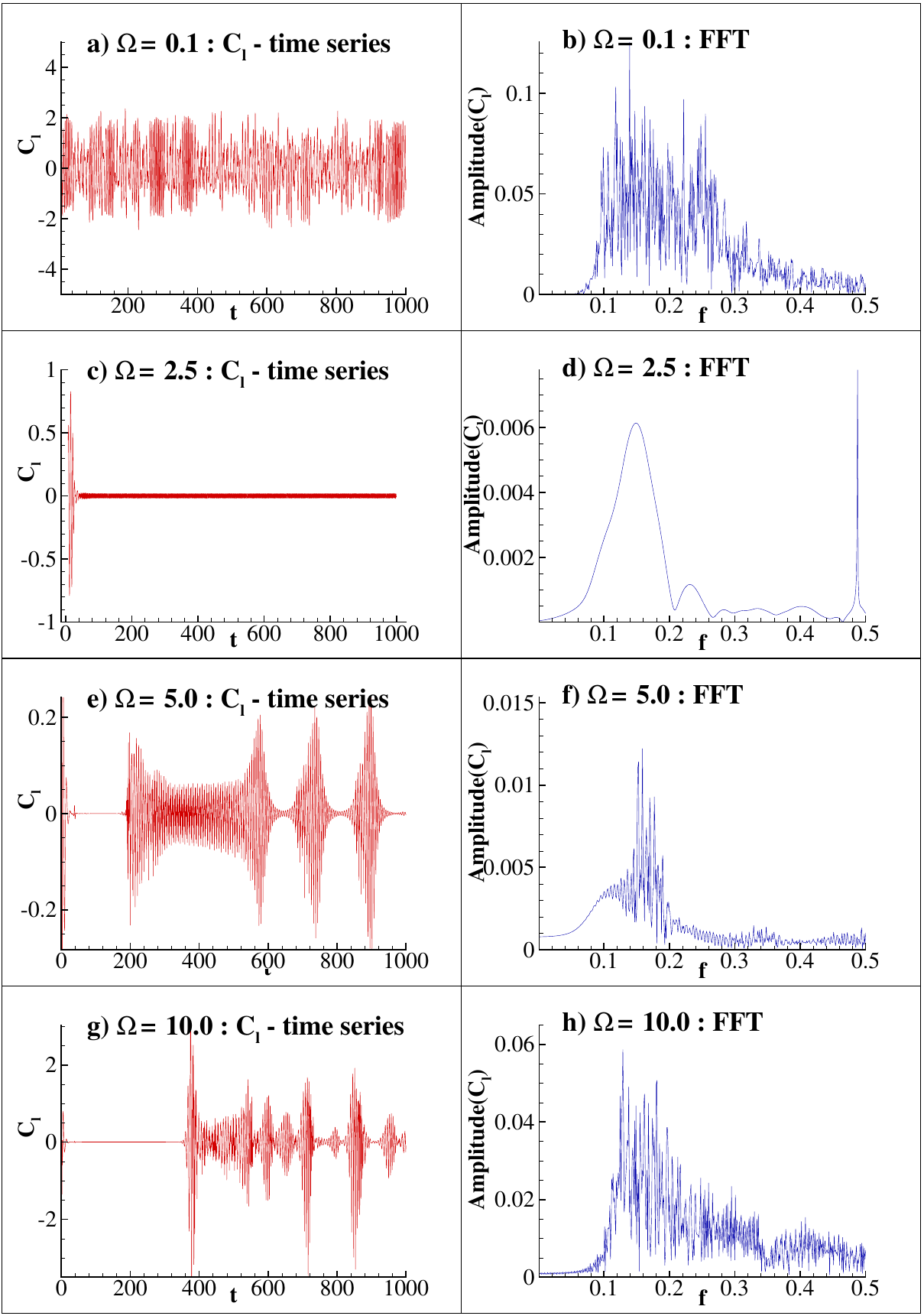}
\caption{Time-series of lift coefficient, $C_l$ for (a) $\Omega = 0.1$, (c) $\Omega = 2.5$, (e) $\Omega = 5.0$, and (g) $\Omega = 10.0$. The corresponding spectra are shown in frames (b), (d), (f), and (h), respectively.}
\label{fig6}
\end{figure}

\begin{figure}[!ht]
\centering
\includegraphics[width=.85\textwidth]{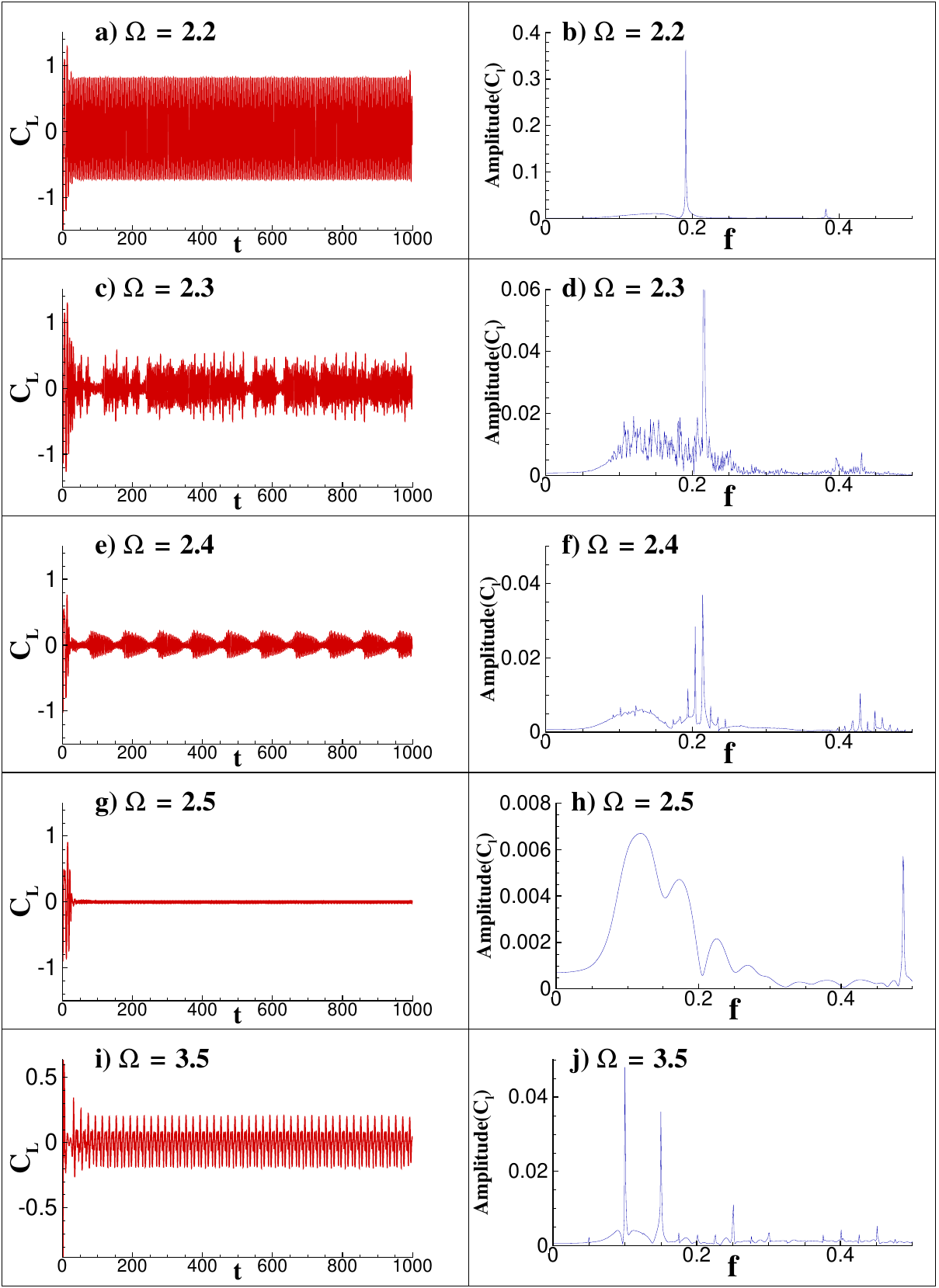}
\caption{Time-series of $C_l$ and its FFT for $\Omega = 2.2$ to 3.5, to explain the threshold $\Omega_t$ beyond which the classical vortex shedding undergoes a fundamental transition.}
\label{fig7}
\end{figure}

\begin{figure}[!ht]
\centering
\includegraphics[width=.65\textwidth]{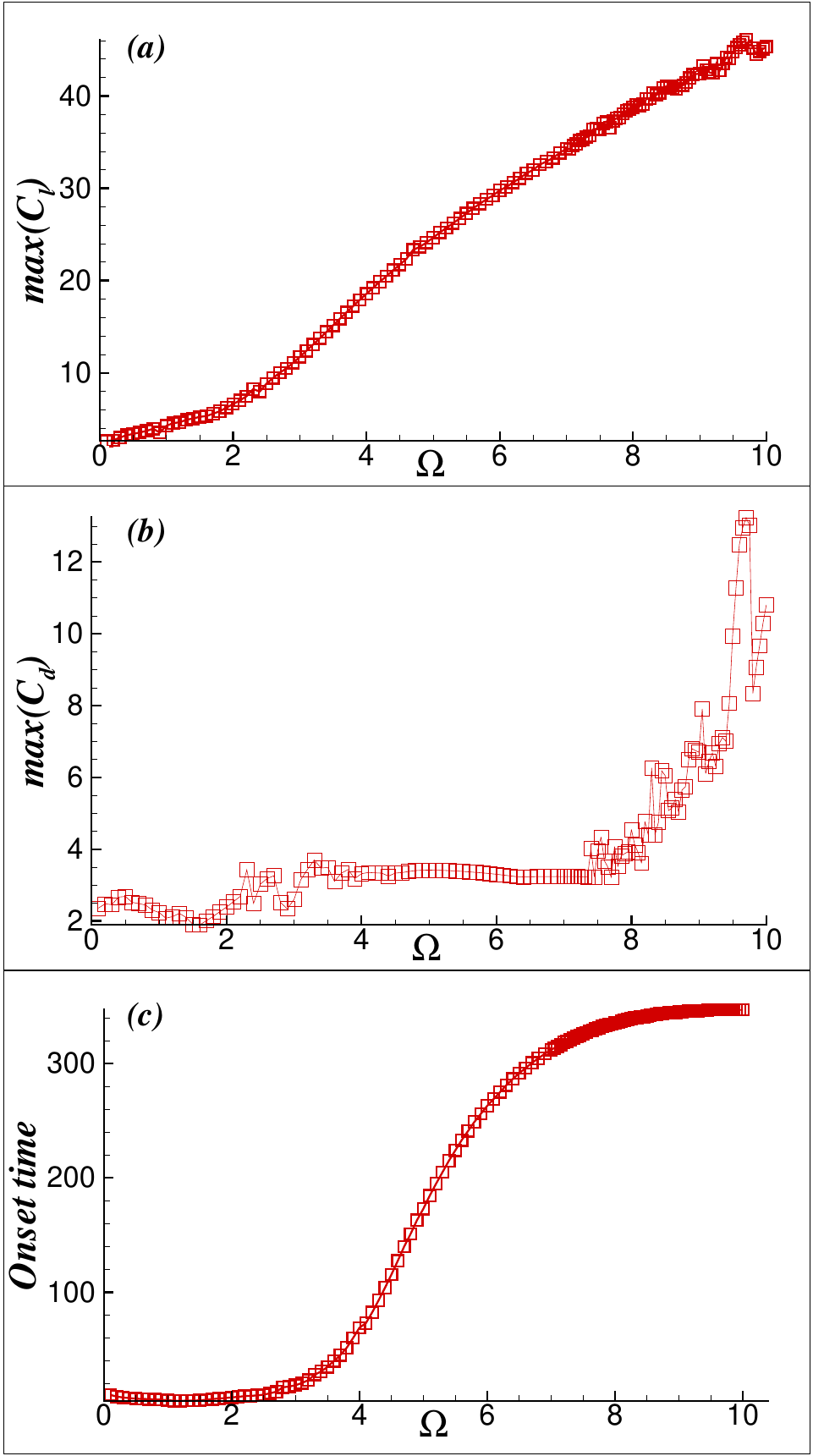}
\caption{Variation of maximum coefficients of (a) lift, (b) drag, and (c) onset time of instability as a function of $\Omega$ for flow past rotating cylinder with $Re_{\infty} = 3800$ and $M_{\infty}$ = 0.1.}
\label{fig8}
\end{figure}

Having established the rotation-induced modifications to the near-body flow, vorticity dynamics, and local compressibility, the discussion is now extended to the corresponding aerodynamic response and the characteristic scales governing the observed unsteadiness. Figures \ref{fig6}–\ref{fig8} examine the temporal evolution of the aerodynamic force coefficients and identify the dominant spatial and temporal scales associated with the low- and high-rotation regimes. Particular emphasis is placed on distinguishing the conventional wake-driven dynamics at low rotation rates from the strongly rotation-dominated unsteady behaviour observed at higher rotation rates, thereby establishing a connection between the aerodynamic loading and the underlying scale-selection mechanisms.


Figure \ref{fig6} presents the temporal evolution of the lift coefficient, \(C_l\), together with the corresponding fast Fourier transform (FFT) spectra for rotation rates, \(\Omega=0.1\), \(2.5\), \(5.0\), and \(10.0\). For the lowest rotation rate, \(\Omega=0.1\) shown in Fig. \ref{fig6}(a), the time-series of $C_l$ exhibits persistent fluctuations throughout the observation interval. The associated spectrum in Fig. \ref{fig6}(b) contains energy distributed over a relatively broad range of frequencies, predominantly between \(f\approx0.1\) and \(0.3\), rather than being concentrated at a single sharply defined frequency. This behaviour indicates that the aerodynamic loading is associated with an unsteady wake containing multiple interacting temporal scales. At low rotation rates, the flow remains connected to the vortex-shedding regime, for which the alternating formation and release of vortices produces oscillatory lift forces. Experimental PIV measurements by \citet{lam2009vortex} showed that, for \(Re=3600\)–\(5000\), vortex shedding persists at low rotation rates, with the wake retaining similarities to the classical von Kármán vortex street. The present \(\Omega=0.1\) case lies well within this low-rotation regime, however, the comparatively broadband spectrum indicates that, for \(Re_\infty=3800\), the instantaneous lift response contains a richer temporal content than a purely single-frequency periodic oscillation.

A fundamentally different response is obtained for \(\Omega=2.5\) depicted in Fig. \ref{fig6}(c). Following an initial transient, the lift fluctuations rapidly decay and \(C_l\) approaches an essentially time-independent state. This strong attenuation of the unsteady aerodynamic loading indicates suppression of the conventional wake instability. The absence of sustained large-amplitude oscillations in the time series is consistent with the progressive suppression of vortex shedding by increasing cylinder rotation reported in previous numerical and experimental investigations \cite{mittal2003flow, lam2009vortex}. Badr {\it et al.} \cite{badr1990unsteady} also reported periodic vortex evolution and shedding at lower rotation rates, while increasing rotation progressively suppressed unsteady flow while investigating unsteady rotating-cylinder flow over Reynolds numbers of \(10^3\) and \(10^4\) and rotation ratios between approximately \(0.5\) and \(3\). The corresponding FFT in Fig. \ref{fig6}(d) contains finite spectral components, but the time-domain signal demonstrates that these are not associated with a sustained limit-cycle oscillation over the entire observation interval. Instead, much of the spectral content is attributable to the initial transient and its subsequent decay. 

The \(\Omega=5\) case shown in Fig. \ref{fig6}(e) reveals the re-emergence of unsteadiness, but with a character distinctly different from the low-rotation wake oscillation. The $C_l$ remains relatively small for extended intervals and is interrupted by localized bursts of oscillatory activity. These intermittent events produce pronounced packets of lift fluctuations rather than a continuously maintained periodic limit cycle. Correspondingly, the FFT in Fig. \ref{fig6}(f) shows a broad distribution of lift force, with enhanced spectral activity concentrated primarily around \(f\approx0.1\)–\(0.2\), together with a cluster of sharper components. This indicates the coexistence of multiple dynamically relevant time scales. Such behaviour is particularly significant because it demonstrates that suppression of the vortex street does not necessarily imply that the flow remains permanently steady at higher rotation rates. This closely follows high-rotation temporal instability \cite{sengupta2003temporal} in the lift and drag of a rapidly rotating and translating cylinder at supercritical rotation rates. Their simulations, performed for conditions including \(Re=3800\), showed irregular temporal instability and established that the high-rotation Magnus–Robins flow possesses an instability mechanism distinct from traditional vortex shedding. The intermittent lift bursts observed at \(\Omega=5\) are therefore more appropriately associated with the emergence of this rotation-dominated temporal instability than with a simple reappearance of the classical von Kármán shedding process.

At the highest rotation rate, \(\Omega=10\), shown in Fig. \ref{fig6}(g), the lift response exhibits an even more pronounced intermittent character. After an extended interval of negligible fluctuations, large-amplitude events emerge abruptly and are followed by alternating periods of intense and relatively weak aerodynamic response. The corresponding spectrum in Fig. \ref{fig6}(h) possesses substantial energy over a broad frequency range, with its strongest components occurring predominantly at low and intermediate frequencies rather than collapsing onto a single dominant spectral peak. The broadband nature of the spectrum is consistent with a temporally complex, multiscale flow response. 

It is worth noting the distinction between the \(\Omega=0.1\) and \(\Omega=10\) cases although both cases exhibit broadband spectral signatures and significant lift fluctuations. At \(\Omega=0.1\), the unsteadiness originates primarily from the wake dynamics associated with the separated shear layers and vortex shedding. At \(\Omega=10\), by contrast, the conventional wake has already undergone substantial rotation-induced reorganization, and the unsteadiness is generated within a strongly rotation-dominated near-body flow. The high-rotation fluctuations are therefore associated with the Magnus–Robins instability \cite{sundaram2021multiscale, sengupta2025bifurcation} and the nonlinear interaction of the rotationally generated vorticity with the surrounding flow.


Figure \ref{fig7} examines the transition between the stabilized state observed at \(\Omega=2.5\) in Fig. \ref{fig6} and the re-emergence of unsteadiness at higher rotation rates by presenting the lift-coefficient time histories and their corresponding FFT spectra for \(\Omega=2.2\), \(2.3\), \(2.4\), \(2.5\), and \(3.5\). For \(\Omega=2.2\) shown in Fig. \ref{fig7}(a,b), the lift coefficient exhibits sustained, high-amplitude oscillations after the initial transient, with fluctuations of the order of \(O(1)\). The corresponding FFT is dominated by a sharp spectral peak near \(f\approx0.19\), whose amplitude is approximately \(0.35\). The concentration of spectral energy in a single dominant mode indicates a highly coherent and strongly periodic aerodynamic response. Thus, although cylinder rotation has substantially modified the traditional wake viewed in a low-rotation regime, the flow at \(\Omega=2.2\) still supports a well-defined instability with a dominant temporal scale.

A marked reduction in the strength of the instability occurs when the rotation rate is increased slightly to \(\Omega=2.3\) in Fig. \ref{fig7}(c,d). Following the initial transient, the amplitude of the $C_l$ fluctuations decreases substantially, and the subsequent signal consists of comparatively weak and irregular oscillations. This reduction is reflected clearly in the FFT spectrum. Although a dominant component remains near \(f\approx0.21\), its amplitude decreases to approximately \(0.06\), nearly an order of magnitude smaller than that at \(\Omega=2.2\). Moreover, the surrounding frequencies have non-negligible $C_l$ amplitude, indicating a loss of the strong single-frequency coherence observed at \(\Omega=2.2\). The transition from \(\Omega=2.2\) to \(2.3\) represents a substantial weakening of the instability rather than merely a change in its dominant frequency.

At \(\Omega=2.4\), shown in Fig. \ref{fig7}(e,f), \(C_l\) exhibits a sequence of intermittent wave packets separated by intervals of relatively weak fluctuations instead of sustained oscillations of approximately constant amplitude. The amplitude modulation in the time series suggests that the flow is approaching a a weakly unsteady/quiescent aerodynamic state but continues to support transient or intermittently amplified disturbances. This behaviour is also evident in the FFT spectrum, where the dominant spectral amplitudes are reduced further to approximately \(0.03\)–\(0.04\). In contrast to the strongly dominant single peak at \(\Omega=2.2\), the spectral amplitude at \(\Omega=2.4\) is distributed among several nearby frequencies. The coexistence of multiple peaks and the amplitude-modulated time signal suggest competition between closely spaced temporal scales and incomplete suppression of the underlying instability.

For \(\Omega=2.5\) shown in Fig. \ref{fig7}(g,h), $C_l$ rapidly approaches an essentially constant value after the initial transient, and remains nearly invariant over the remainder of the time interval. The reduction in aerodynamic fluctuations is dramatic when compared with \(\Omega=2.2\), The FFT amplitudes decrease from approximately \(0.35\) at \(\Omega=2.2\) to less than \(0.007\) at \(\Omega=2.5\). Thus, the dominant unsteady loading has been attenuated by almost two orders of magnitude. The remaining low-frequency spectral content has a very small amplitude and should not be interpreted as evidence of a sustained large-scale instability. Instead, it is consistent with the finite transient preceding the establishment of the a weakly unsteady/quiescent aerodynamic regime. Intermediate rotation rates \(2.6\leq\Omega\leq3.4\) are not shown as their $C_l$ histories remain qualitatively similar to that obtained at \(\Omega=2.5\). This indicates that the suppression is not confined to a single critical value of rotation rate. Rather, there is a finite stable or weakly unsteady interval extending beyond \(\Omega=2.5\). This is important because it demonstrates that the suppression of the low-rotation instability persists over a range of rotational forcing.

The behaviour changes again at \(\Omega=3.5\), as depicted in Fig. \ref{fig7}(i,j), where sustained lift fluctuations reappear. However, the character of this unsteadiness differs substantially from that at \(\Omega=2.2\). The amplitude of the time-domain oscillations is considerably smaller than in the \(\Omega=2.2\) case, while the FFT no longer exhibits one overwhelmingly dominant spectral component. Instead, two pronounced peaks, with amplitudes of approximately \(0.05\) and \(0.035\), occur at distinct frequencies, accompanied by weaker higher-frequency components. The re-emergence of multiple energetic frequencies indicates that the instability appearing at \(\Omega=3.5\) is not simply a continuation of the low-rotation oscillatory mode.

Effect of rotation on aerodynamic instability is strongly non-monotonic. Increasing rotation rate initially suppresses the instability responsible for the large-amplitude lift oscillations, producing a finite interval of aerodynamic stabilization. Further increases in rotation subsequently permit the development of a different unsteady state. This provides a more detailed interpretation of the behaviour observed in Fig. \ref{fig6}. The transition between the low-rotation and high-rotation regimes is mediated by a distinct window in which the dominant lift fluctuations collapse in amplitude before a new, rotation-dominated instability emerges.


Figure \ref{fig8} summarizes dependence of the maximum lift coefficient, \(\max(C_l)\), maximum drag coefficient, \(\max(C_d)\), and the onset time of instability on the rotation rate, \(\Omega\). The onset time is defined as the time beyond which the $C_l$ evolution shows a growth after the initial transients. Together, these quantities provide a global view of how increasing rotational forcing modifies both the aerodynamic loading and the temporal stability of the flow. As shown in Fig. \ref{fig8}(a), \(\max(C_l)\) increases monotonically with \(\Omega\), reflecting the progressive strengthening of the Magnus-Robins effect. At low rotation rates, the increase is relatively gradual because the rotational velocity remains comparable to or smaller than the free-stream velocity. Beyond \(\Omega\approx2\), however, the growth becomes substantially more pronounced as rotation increasingly dominates the near-body momentum distribution and produces a stronger asymmetry in the surface pressure field. The approximately monotonic increase in maximum $C_l$ is consistent with the classical understanding of the Magnus-Robins force and with previous experimental and numerical studies of rotating cylinders \cite{tokumaru1991rotary, pralits2013three}.

In contrast, the variation of \(\max(C_d)\) in Fig. \ref{fig8}(b) is distinctly non-monotonic. At low and intermediate rotation rates, the maximum drag remains relatively moderate, with only limited variation. This behaviour is associated with the progressive suppression and reorganization of the conventional separated wake by rotation. At higher rotation rates, particularly beyond \(\Omega\approx7\), \(\max(C_d)\) increases rapidly and exhibits substantial fluctuations from one rotation rate to another. This sharp increase is consistent with the emergence of strongly unsteady, rotation-dominated flow dynamics \cite{kumar2026supervised}.

The onset time shown in Fig. \ref{fig8}(c) provides further insight into this transition. The instability appears rapidly at low rotation rates, but the onset time increases sharply with increasing \(\Omega\), reaching values exceeding \(300\) at high rotation rates. The delay indicates that rotational forcing initially stabilizes the wake-induced instability by suppressing the alternating vortex-shedding mechanism. However, the stabilization does not imply the complete elimination of instability. Instead, as demonstrated by the lift force histories and spectral results in Figs. \ref{fig6} and \ref{fig7}, a different instability mechanism eventually develops at higher rotation rates \cite{sundaram2021multiscale, suman2022novel}. The onset time subsequently approaches an approximately saturated value, suggesting that the high-rotation instability possesses a characteristic growth process that is distinct from the wake-induced instability at lower rotation rates.

\subsection{Application of vorticity-based proper orthogonal decomposition}

\begin{figure}[!ht]
\centering
\includegraphics[width=.9\textwidth]{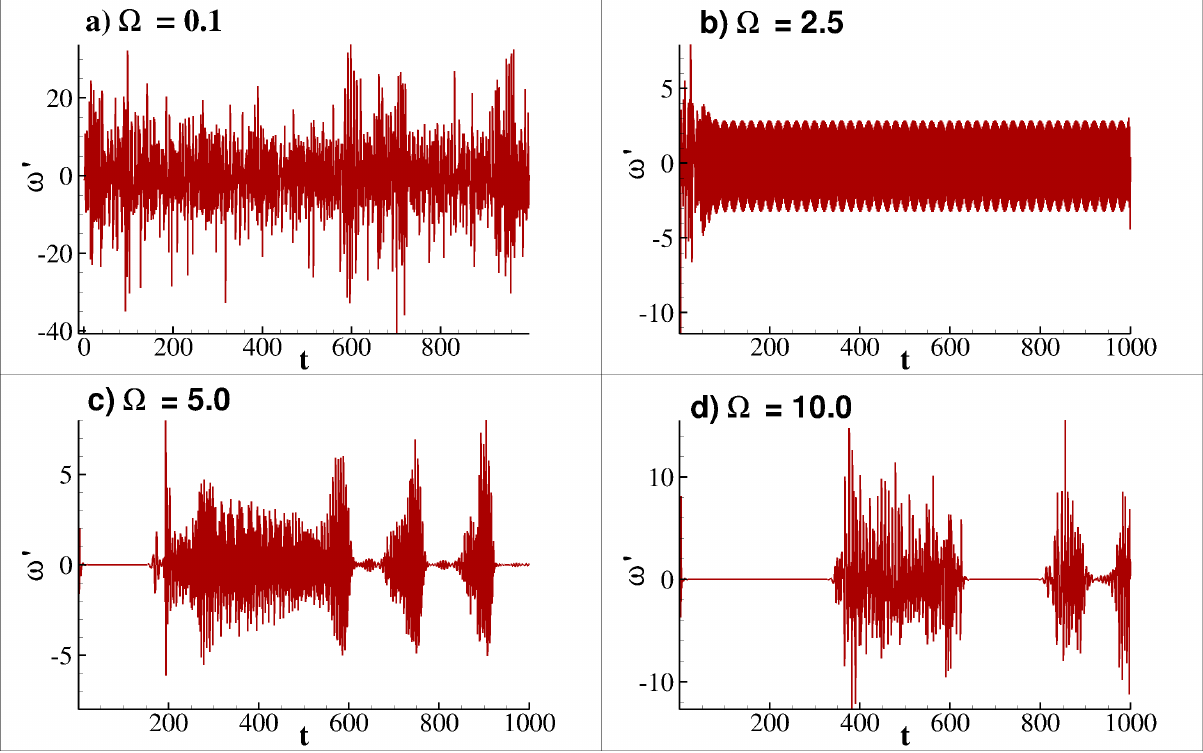}
\caption{Evolution of disturbance vorticity is shown for (a) $\Omega = 0.1$, (b) $\Omega = 2.5$, (c) $\Omega = 5$, and (d) $\Omega = 10$, to form the basis of vorticity-based POD. The probe location is considered at the centre of the primary vortex in the wake of the cylinder.}
\label{fig9}
\end{figure}

The preceding analysis has demonstrated that the flow undergoes substantial changes in its temporal behaviour and instability characteristics as the rotation rate is increased. To identify the dominant coherent structures underlying these dynamics and quantify their relative energetic contributions, a proper orthogonal decomposition (POD) is subsequently performed using the disturbance vorticity ($\omega'$) field. Here, $\omega'$ is defined as $\omega' = \omega - \overline{\omega}$, where $\omega$ is the instantaneous vorticity and $\overline{\omega}$ is the time-averaged value.  Figure \ref{fig9} presents the temporal evolution of $\omega'$ for the four representative rotation rates, \(\Omega=0.1\), \(2.5\), \(5\), and \(10\), which form the snapshot database for the subsequent POD analysis. The decomposition is then used to examine how the dominant spatial modes and their associated temporal behaviour evolve from the low-rotation regime to the strongly rotation-dominated flow.

In Fig. \ref{fig9}, the $\omega'$ signal is probed at the core of the primary vortex in the wake of the cylinder, i.e., in a region of intense vortical activity. For \(\Omega=0.1\), Fig. \ref{fig9}(a) shows persistent fluctuations throughout the sampling interval, with the largest excursions ranging approximately from \(-40\) to \(+30\). The comparatively large amplitude reflects the sustained vortical activity associated with the low-rotation wake. At \(\Omega=2.5\) shown in Fig. \ref{fig9}(b), the initial transient is followed by persistent oscillations of substantially smaller amplitude, predominantly within approximately \(\omega'\approx\pm3\), although the initial fluctuations reach considerably larger values.

This observation is important when compared with the $C_l$ evolution shown in Fig. \ref{fig6}. Although the \(C_l\) signal at \(\Omega=2.5\) indicates an essentially stabilized aerodynamic response, the vorticity evolution demonstrates that localized vortical fluctuations persist within the wake. It should be noted that \(C_l\) represents the integrated aerodynamic lift evaluated from the pressure and viscous stresses on the cylinder surface, whereas \(\omega'\) is measured locally at the core of the primary wake vortex. Consequently, relatively weak but persistent vortical motion in the wake need not produce appreciable fluctuations in the net lift acting on the cylinder.

At \(\Omega=5.0\) shown in Fig. \ref{fig9}(c), $\omega'$ exhibits a distinctly intermittent character. Extended intervals of relatively weak fluctuations are interrupted by localized bursts of intense vortical activity, with excursions reaching approximately \(\omega'\approx\pm6\)–\(8\). The intermittent nature of the signal is consistent with the re-emergence of the high-rotation instability identified from $C_l$ response in Fig. \ref{fig6}. The most pronounced intermittency is observed for \(\Omega=10.0\) in Fig. \ref{fig9}(d). The signal contains extended periods of nearly negligible $\omega'$ separated by intense bursts, with peak amplitudes exceeding \(\pm10\) and approaching approximately \(\pm15\). This behaviour indicates that the high-rotation flow is characterized by strongly localized and temporally intermittent vortical events rather than continuously sustained oscillations of uniform amplitude.


For the POD analysis, the space- and time-dependent disturbance vorticity field, \(\omega'(\mathbf{X},t)\), is considered over the temporal interval \(400\leq t\leq800\), which encompasses the growth and subsequent decay of the instability for all $\Omega$, as illustrated in Fig. \ref{fig9}. A total of \(M=801\) snapshots of the vorticity field in the \((x,y)\)-plane are employed to construct the snapshot ensemble. The disturbance field is represented through the POD expansion as

\begin{equation}
\omega'(\mathbf{X},t)=\sum_{m=1}^{M}a_m(t)\phi_m(\mathbf{X}),
\end{equation}

\noindent where \(\phi_m(\mathbf{X})\) denotes the \(m^{\mathrm{th}}\) spatial POD mode and \(a_m(t)\) is its corresponding temporal amplitude. The POD modes are obtained from the eigen decomposition of the two-point spatial correlation operator constructed from the snapshot ensemble. For the discrete snapshots at times \(t_i\) and \(t_j\), the correlation matrix is defined as
\vspace{-3mm}
$$ R_{ij}=\frac{1}{M}\int_S \omega'(\mathbf{X},t_i)\, \omega'(\mathbf{X},t_j)\,dS, $$

\noindent where \(i,j=1,2,\ldots,M\), and \(S\) denotes the spatial domain over which the decomposition is performed. The resulting eigenvectors define the POD basis, with the associated eigenvalues providing a measure of the relative energy captured by each mode. The spatial modes are therefore often referred to as eigenmodes.

Once the orthogonal POD basis has been determined over the selected time interval, the temporal amplitude associated with the \(i^{\mathrm{th}}\) mode is obtained by projection of the instantaneous disturbance field onto the corresponding spatial mode,
\vspace{-2mm}
$$ a_i(t)= \frac{\displaystyle\int_S \omega'(\mathbf{X},t)\phi_i(\mathbf{X})\,dS} {\displaystyle\int_S \phi_i^2(\mathbf{X})\,dS}. $$

Thus, each POD mode represents an optimal spatial structure extracted from ensemble of vorticity snapshots, while its associated amplitude describes temporal evolution of that structure. The decomposition consequently provides a compact representation of the unsteady flow field by ranking the coherent structures according to their energetic contributions. This enables the dominant spatial patterns and their associated temporal dynamics to be identified while substantially reducing the dimensionality of the original high-dimensional dataset.

\begin{figure}[!ht]
\centering
\includegraphics[width=.8\textwidth]{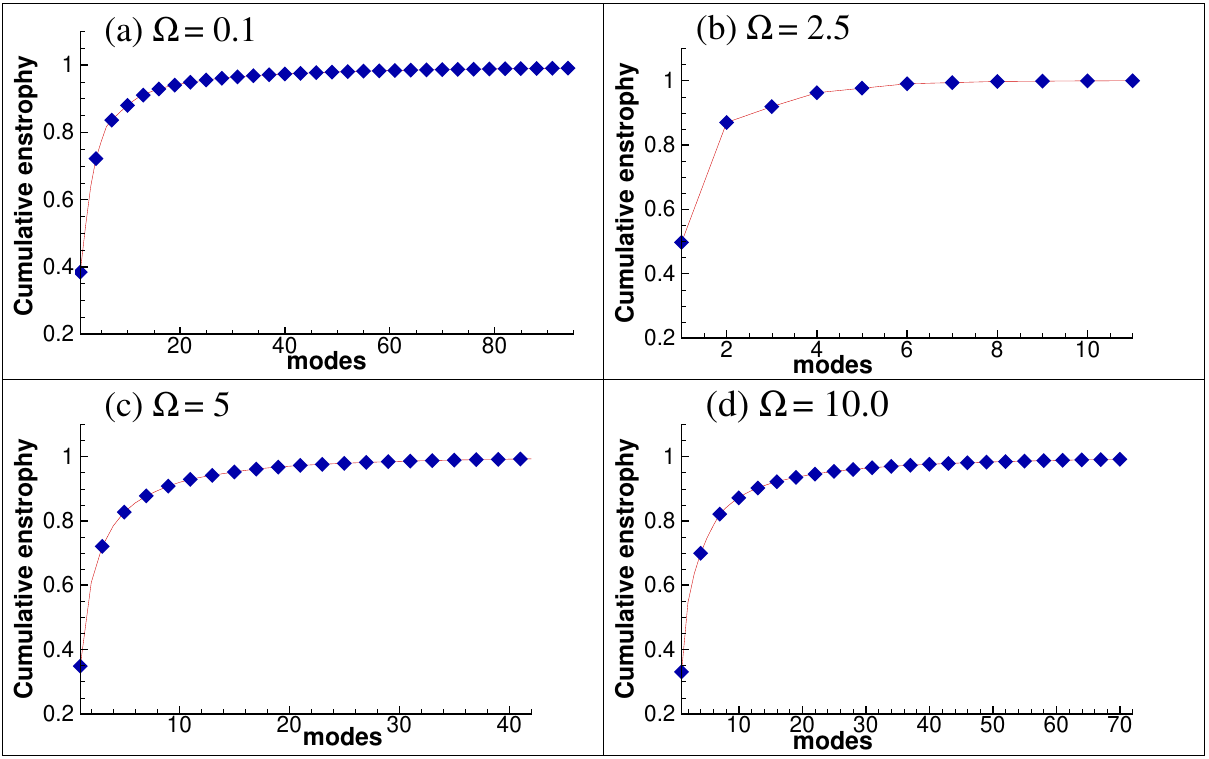}
\caption{Cumulative enstrophy captured by POD modes for $\omega'$ at \(\Omega=0.1\), \(2.5\), \(5.0\), and \(10.0\) is shown as a function of number of modes. We use number of modes required to capture \(99\%\) of the total enstrophy as the threshold for reconstruction.}
\label{fig10}
\end{figure}

Figure \ref{fig10} shows the cumulative enstrophy, defined as the spatially integrated squared vorticity (measure of intensity and distribution of rotational motion in the flow), captured by the POD modes. The results reveal substantial differences in effective dimensionality of the vorticity dynamics across the four rotation rates.

For \(\Omega=0.1\), approximately 90 eigenmodes are required to capture \(99\%\) of the total enstrophy, indicating that the low-rotation wake contains a broad range of energetically relevant vortical structures. This is consistent with the persistent, large-amplitude vorticity fluctuations observed in Fig. \ref{fig9}, reflecting continuously evolving and spatially distributed wake dynamics. In sharp contrast, only 6 modes are required for \(\Omega=2.5\). The rapid convergence of the cumulative enstrophy indicates that the vortical dynamics are dominated by a small number of coherent structures. Although Fig. \ref{fig9} shows persistent localized vorticity fluctuations at the probe location, their comparatively weak and organized character results in a highly low-dimensional representation of the flow. At \(\Omega=5\), the number of modes required increases to 37, indicating the re-emergence of a broader range of dynamically relevant scales. This is consistent with the intermittent bursts observed in the vorticity time series of Fig. \ref{fig9}, which redistribute the enstrophy among a larger number of POD modes. At the highest rotation rate, \(\Omega=10\), to capture \(99\%\) of the total enstrophy, 66 eigenmodes are required, indicating that $\omega'$ field has a higher effective modal dimensionality. The intense intermittent events observed in Fig. \ref{fig9} therefore involve a substantially wider range of coherent structures than the \(\Omega=2.5\) case.

\begin{figure}[!ht]
\centering
\includegraphics[width=.85\textwidth]{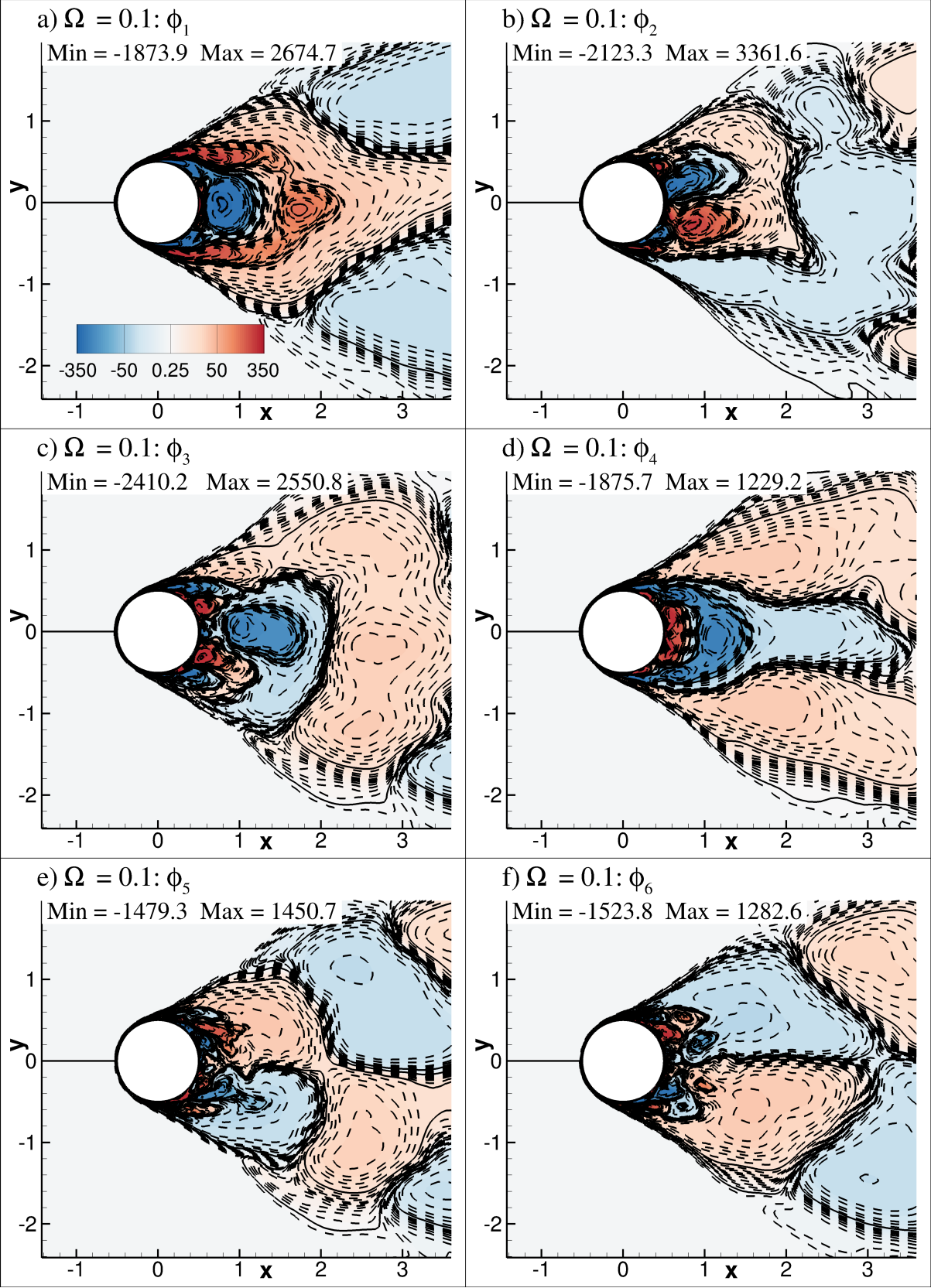}
\caption{Contour plots of first six spatial/eigenfunction modes ($\phi_1$ - $\phi_6$) from POD applied to $\Omega = 0.1$ case are shown in frames (a)-(f). The POD is evaluated over $400 \le t \le 800$ of Fig. \ref{fig9}(a).}
\label{fig11}
\end{figure}

Figure \ref{fig11} presents first six eigenmodes of $\omega'$ for \(\Omega=0.1\). The modes reveal the hierarchy of coherent vortical structures contributing to the strongly unsteady wake identified in Fig. \ref{fig9}. The first two modes, \(\phi_1\) and \(\phi_2\) shown in Figs. \ref{fig11}(a)-(b), exhibit the most regular and clearly organized structures. Both contain concentrated regions of positive and negative vorticity in the near wake, with broadly similar spatial patterns displaced relative to one another. Such modal pairing is characteristic of convecting and approximately periodic wake dynamics, where two POD modes can represent different phases of the same coherent oscillatory process \cite{corrochano2026coherent}. The regularity of these modes is consistent with the persistent large-amplitude fluctuations observed in Fig. \ref{fig9}(a). Similar pairing of the leading POD modes has been widely observed in canonical cylinder wakes, where the first modal pair represents the dominant vortex-shedding dynamics \cite{sengupta2015enstrophy}.

The third mode, \(\phi_3\) in Fig. \ref{fig11}(c), displays the richest spatial organization among the first six modes, particularly in the immediate wake. Multiple regions of alternating sign are present behind the cylinder, together with a large-scale downstream structure, indicating that this mode contains contributions from several interacting wake scales rather than a single simple vortical pattern. In this sense, \(\phi_3\) provides the clearest indication of the multiscale nature of the low-rotation wake. Modes \(\phi_4\)–\(\phi_6\) illustrated in Figs. \ref{fig11}(d)-(f) exhibit a combination of localized near-cylinder structures and larger wake-scale regions extending downstream. In particular, \(\phi_4\) is dominated by an elongated region of negative vorticity along the central wake, accompanied by comparatively broader positive-vorticity regions away from the centreline. By contrast, \(\phi_5\) and \(\phi_6\) contain increasingly localized alternating-sign structures close to the cylinder, superimposed on larger-scale downstream features. These higher-order modes therefore appear to represent finer spatial corrections to the dominant large-scale organization captured by the leading modes. Presence of increasingly complex structures in modes \(\phi_3\)–\(\phi_6\), together with the requirement of 90 modes to capture \(99\%\) of the total enstrophy shown in Fig. \ref{fig10}, demonstrates that the wake at \(Re_\infty=3800\) cannot be represented solely by a single modal pair. Rather, the dominant coherent structures coexist with a broad hierarchy of additional vortical scales.

An important feature of Fig. \ref{fig11} is the asymmetry in the spatial concentration of positive and negative vorticity. The strongest localized positive and negative regions occur predominantly in the near wake, where the separated shear layers interact and roll up into coherent vortical structures. The relative dominance of one sign within individual modes should not, however, be interpreted as a net physical excess of vorticity of that sign in the instantaneous flow. Each POD eigenmode represents an orthogonal mathematical structure whose positive and negative regions combine through the time-dependent modal coefficient, $a_m(t)$ to reconstruct the physical disturbance field.

\begin{figure}[!ht]
\centering
\includegraphics[width=.85\textwidth]{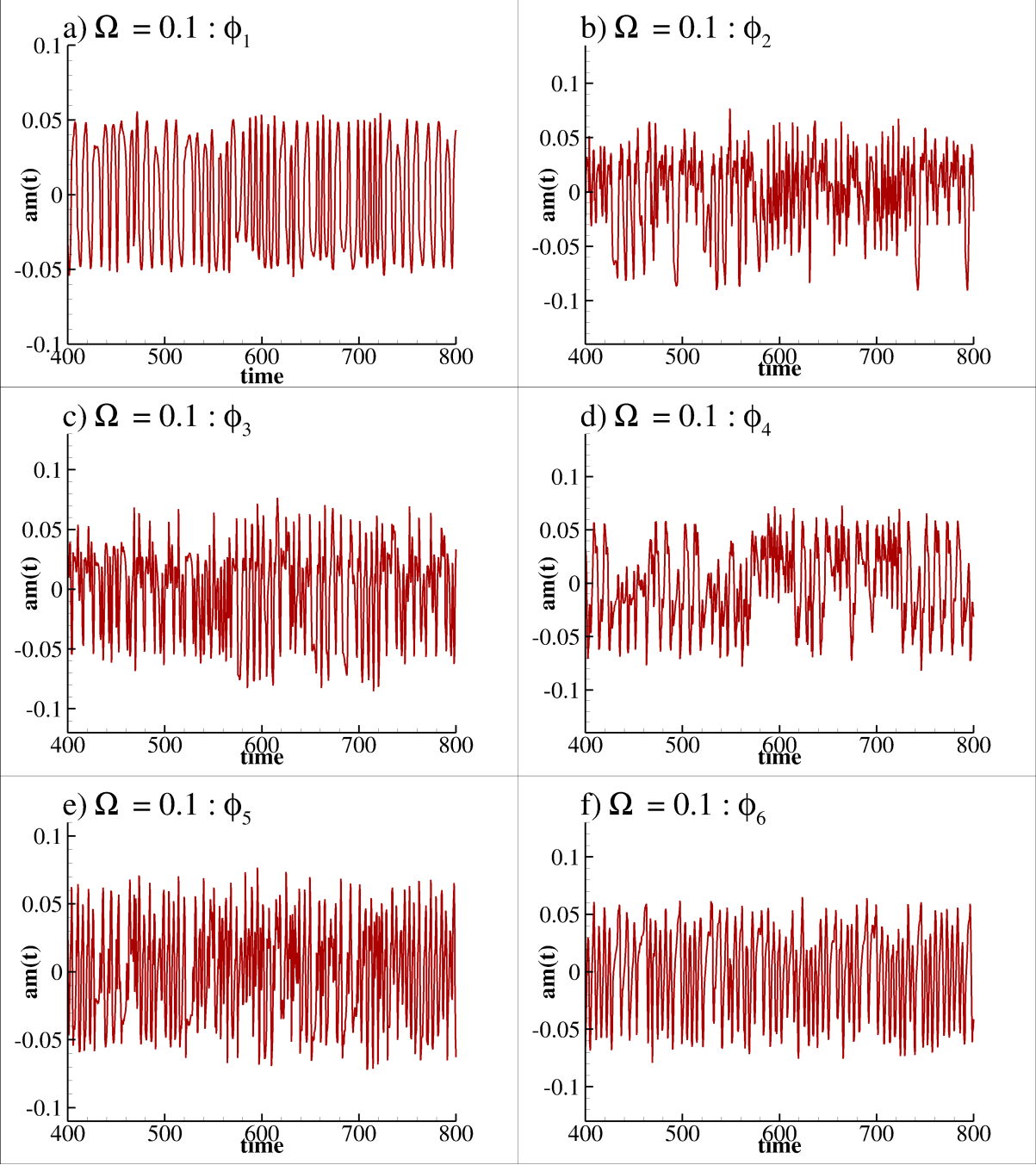}
\caption{Time-dependent amplitude functions of first six POD modes in frames (a)-(f), obtained for $\Omega = 0.1$ case, over time interval, $400 \le t \le 800$ in Fig. \ref{fig9}(a).}
\label{fig12}
\end{figure}

Figure \ref{fig12} presents the temporal evolution of the modal coefficients, \(a_m(t)\), associated with the first six POD eigenmodes for \(\Omega=0.1\). The coefficients quantify the instantaneous contribution of each spatial mode to the reconstruction of $\omega'$, providing temporal counterpart to the spatial structures shown in Fig. \ref{fig11}. The first modal coefficient, \(a_1(t)\) shown in Fig. \ref{fig12}(a) exhibits the clearest temporal organization among the six modes. It is characterized by sustained oscillations with a relatively well-defined amplitude, predominantly within approximately \(\pm0.05\). This comparatively regular behaviour is consistent with the coherent wake structures represented by the leading mode. In contrast, \(a_2(t)\) shown in Fig. \ref{fig12}(b) displays substantially more irregular fluctuations, with intermittent excursions approaching \(\pm0.1\). The larger instantaneous excursions of $\phi_2$ indicate that, despite its lower modal rank, the corresponding structure undergoes strong temporal modulation.

The coefficients associated with modes \(3\)–\(6\), shown in Figs. \ref{fig12}(c)-(f), exhibit increasingly irregular temporal behaviour. Their fluctuations remain predominantly within approximately \(\pm0.05\)–\(0.08\), but without the clear periodic organization apparent in \(a_1(t)\). In particular, \(a_3(t)\) and \(a_5(t)\) show strongly fluctuating, broadband-like signals, whereas \(a_4(t)\) and \(a_6(t)\) retain intermittent intervals of more organized oscillatory behaviour. The time histories reinforce the conclusion from Fig. \ref{fig10} that the \(\Omega=0.1\) wake is not governed by a small number of purely periodic structures. Rather, a hierarchy of coherent and less coherent vortical motions contributes to the overall disturbance field.

In a canonical periodic wake, paired POD modes often possess similar temporal amplitudes and exhibit an approximately \(90^\circ\) phase difference, together representing the same convecting vortex-shedding process \cite{cagney2014streamwise, sengupta2015enstrophy}. The present signals, particularly \(a_1(t)\) and \(a_2(t)\), do not show such a simple correspondence. This departure is physically consistent with the more complex wake dynamics at \(Re_\infty=3800\), where the low rotation modifies the symmetry of the separated shear layers and introduces multiple interacting temporal scales.

\begin{figure}[!ht]
\centering
\includegraphics[width=.85\textwidth]{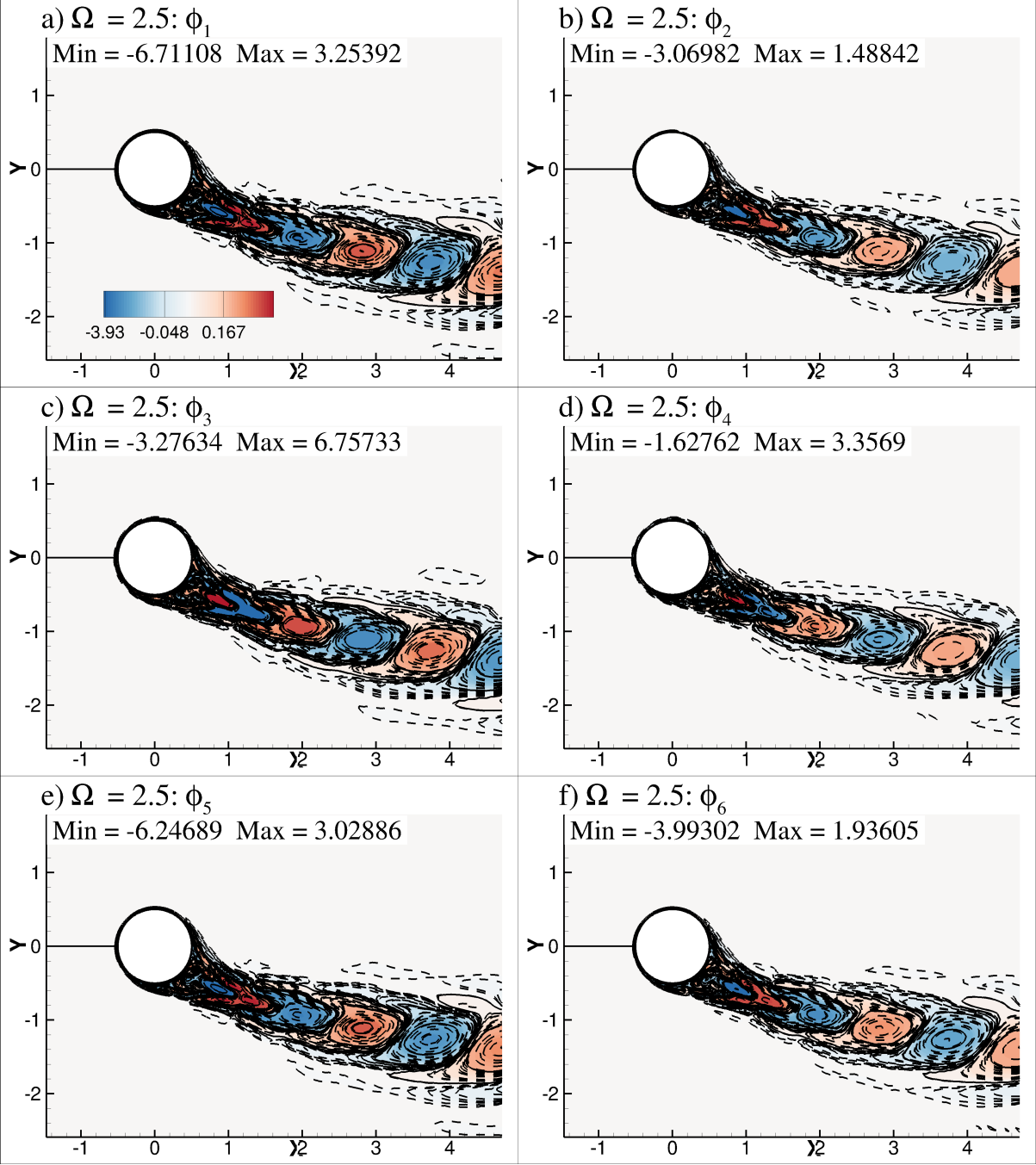}
\caption{Contour plots of first six spatial/eigenfunction modes ($\phi_1$ - $\phi_6$) from POD applied to $\Omega = 2.5$ case are shown in frames (a)-(f). The POD is evaluated over $400 \le t \le 800$ of Fig. \ref{fig9}(b).}
\label{fig13}
\end{figure}

Figure \ref{fig13} presents the first six POD eigenmodes of the disturbance-vorticity field for \(\Omega=2.5\). The most significant feature of these modes is that the residual vortical dynamics are organized along a single, spatially confined, downward-deflected wake trajectory. Rather than occupying a broad region of the computational domain, the energetic disturbance structures remain concentrated within the primary wake, indicating that the dynamics at this rotation rate are strongly constrained by the rotation-modified mean flow. The leading modes exhibit a sequence of alternating-sign vorticity lobes distributed along the deflected wake. For \(\phi_1\) shown in Fig. \ref{fig13}(a), dominant structure consists of succession of relatively large vortical regions extending downstream, whereas \(\phi_2\) shown in Fig. \ref{fig13}(b) reveals closely related spatial organization with a modified distribution of the positive- and negative-vorticity regions. The spatial similarity between these two modes is suggestive of a modal pair associated with the evolution of a coherent convecting disturbance \cite{sengupta2015enstrophy}, although the phase relationship between their corresponding temporal coefficients must be examined to establish this conclusively.

A similar organization is apparent in the subsequent modes. Modes \(\phi_3\) and \(\phi_4\) shown in Figs. \ref{fig13}(c)-(d) contain more closely spaced alternating-sign structures, indicating the representation of comparatively shorter spatial scales within the same wake region. Likewise, \(\phi_5\) and \(\phi_6\) in Figs. \ref{fig13}(e)-(f) remain localized along the primary wake but describe progressively finer spatial variations. Thus, the modal sequence does not indicate the emergence of dynamically independent structures distributed throughout the domain. Instead, the POD resolves a hierarchy of spatial scales embedded within the same rotation-deflected wake.

These results provide an important interpretation of the vorticity signal in Fig. \ref{fig9}(b). At \(\Omega=2.5\), the probe located at the core of the primary wake vortex records persistent but relatively weak fluctuations. Figure \ref{fig13} shows that these fluctuations arise from the temporal evolution of a small number of coherent structures confined to the primary wake. Thus, the weak vorticity oscillations are not indicative of a spatially complex or high-dimensional instability. On the contrary, the spatial organization of the modes, together with the result in Fig. \ref{fig10} that only six modes capture \(99\%\) of the total enstrophy, demonstrates that the residual dynamics at \(\Omega=2.5\) are strongly low-dimensional. The rotation-modified flow at \(\Omega=2.5\) continues to support coherent disturbances within the primary wake, as evident from Fig. \ref{fig9}, but these disturbances are spatially constrained and energetically concentrated into a small number of modes. Consequently, they do not generate the large-scale, spatially evolving dynamics required to produce appreciable fluctuations in the $C_l$ distribution in Fig. \ref{fig6}. The POD therefore identifies \(\Omega=2.5\) not as a completely quiescent state, but as a regime in which the dynamics have collapsed onto a small number of coherent wake structures.

\begin{figure}[!ht]
\centering
\includegraphics[width=.85\textwidth]{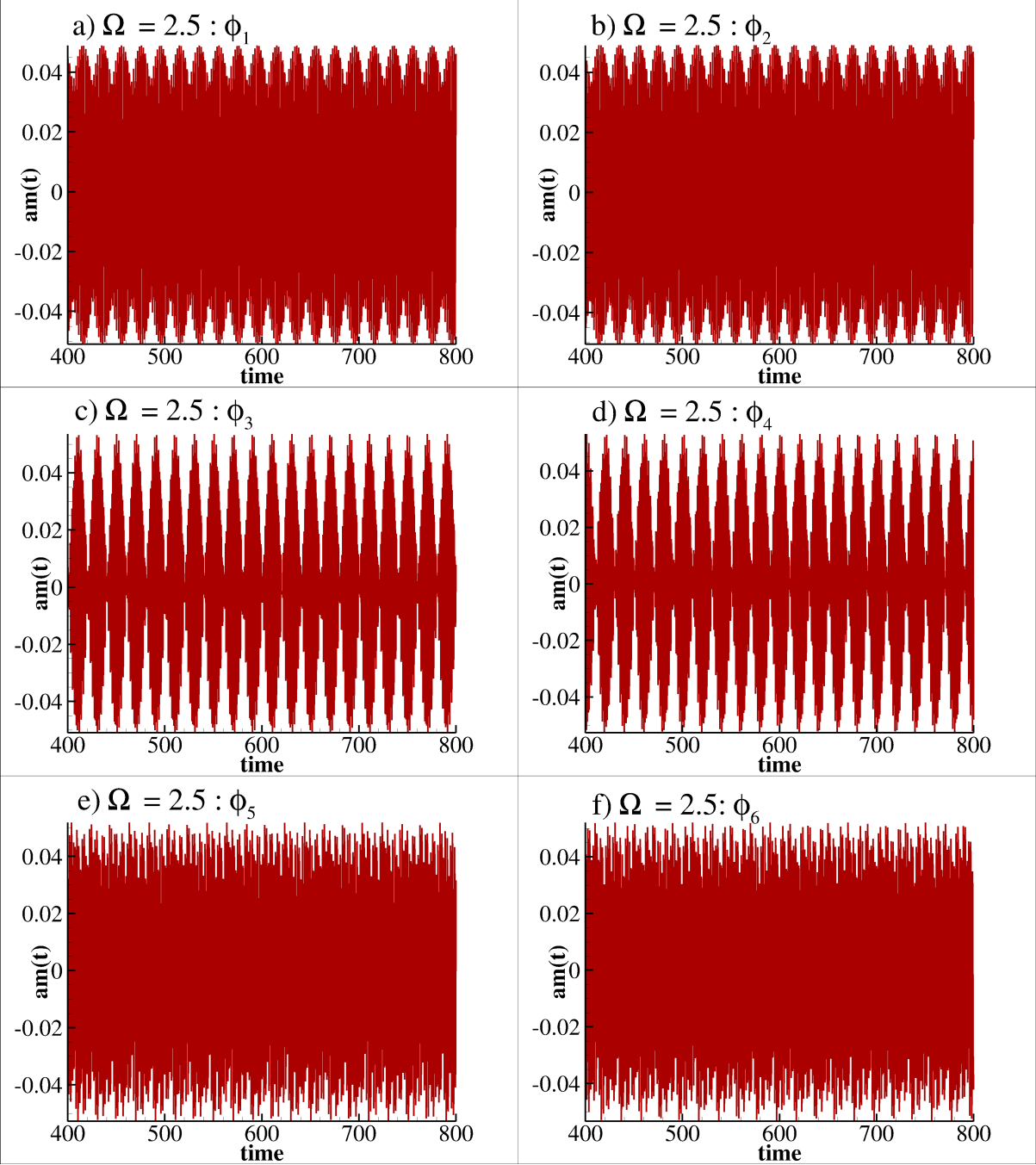}
\caption{Time-dependent amplitude functions of first six POD modes in frames (a)-(f), obtained for $\Omega = 2.5$ case, over time interval, $400 \le t \le 800$ in Fig. \ref{fig9}(b).}
\label{fig14}
\end{figure}

Figure \ref{fig14} shows the temporal coefficients \(a_m(t)\) associated with the first six POD modes for \(\Omega=2.5\). All six coefficients remain bounded within a comparable amplitude range, approximately \(\pm 0.05\), confirming that the residual vortical dynamics identified in Fig. \ref{fig9}(b) are relatively weak and do not exhibit sustained growth over the sampling interval. The temporal evolution of \(a_1(t)\)–\(a_4(t)\) exhibits a more discernible oscillatory organization, with recurrent fluctuations of relatively regular amplitude. In particular, the temporal signatures of the first two modes are broadly similar, suggesting that they may form a complementary modal pair associated with the evolution of the dominant coherent wake structure. Modes \(3\) and \(4\) also display comparatively organized oscillations, whereas \(a_5(t)\) and \(a_6(t)\) are more irregular and contain stronger high-frequency fluctuations.

The absence of large-scale amplitude modulation is consistent with the localized, low-dimensional vortical dynamics observed in Figs. \ref{fig9} and \ref{fig13}. This behaviour also agrees qualitatively with the analysis of rotating-cylinder wakes \cite{pralits2010instability, lam2009vortex}, where increasing rotation substantially disrupted the conventional Kármán shedding process and concentrated the coherent motion into modified large-scale structures. The present results indicate that, at \(\Omega=2.5\), the residual wake fluctuations remain temporally active but are bounded and organized, rather than developing into the large-amplitude global oscillations characteristic of an unstable vortex-shedding state.

\begin{figure}[!ht]
\centering
\includegraphics[width=.85\textwidth]{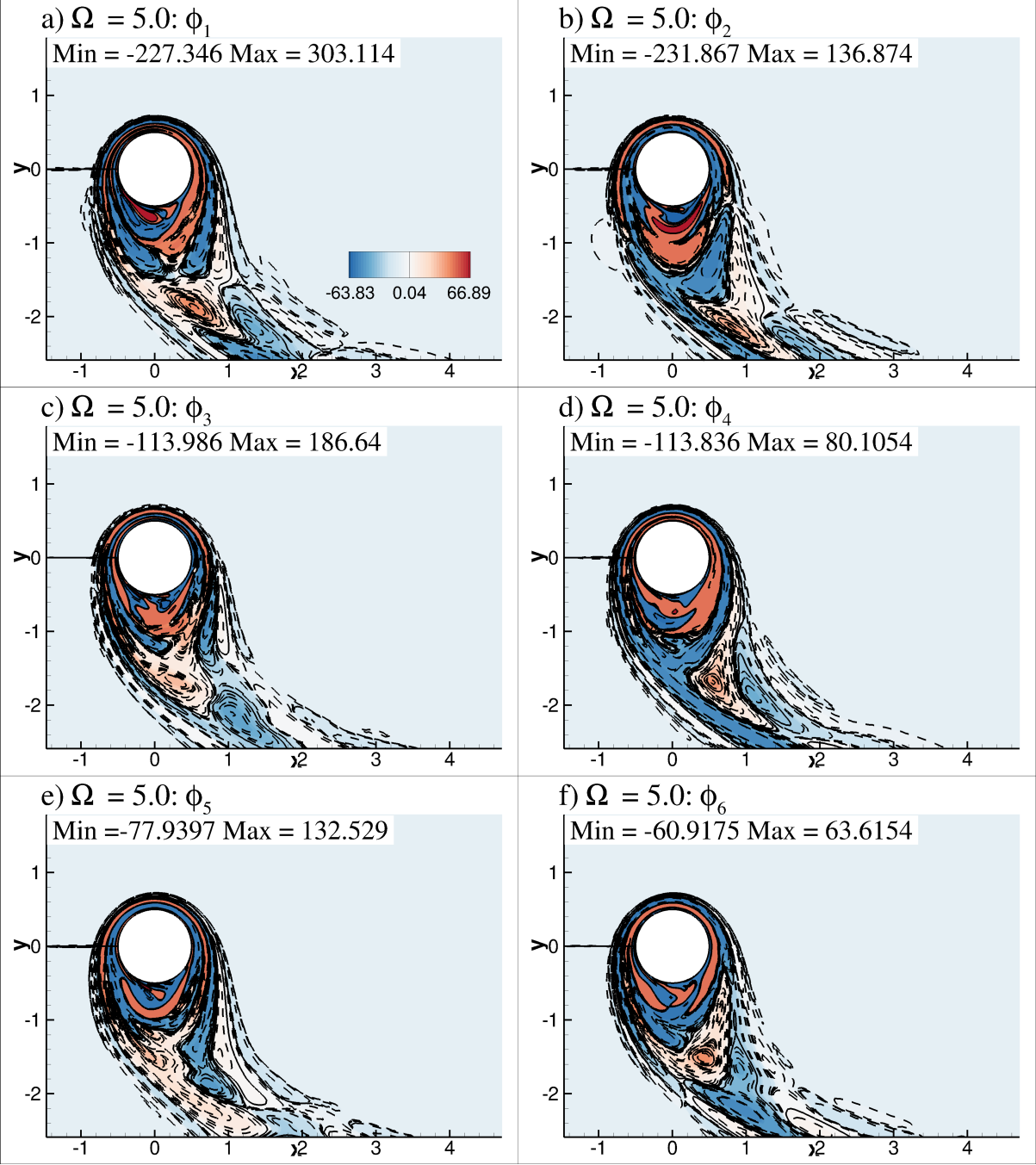}
\caption{Contour plots of first six spatial/eigenfunction modes ($\phi_1$ - $\phi_6$) from POD applied to $\Omega = 5.0$ case are shown in frames (a)-(f). The POD is evaluated over $400 \le t \le 800$ of Fig. \ref{fig9}(c).}
\label{fig15}
\end{figure}

Figure \ref{fig15} presents the first six POD eigenmodes of the disturbance-vorticity field for \(\Omega=5\). Here, the dominant structures are concentrated immediately around the rotating cylinder and extend into a strongly deflected downstream wake. The modes indicate that the energetic dynamics at \(\Omega=5\) involve both near-body rotational structures and their subsequent downstream evolution.

The most distinctive feature is the annular or wrapped vorticity organization surrounding the cylinder. In \(\phi_1\) and \(\phi_2\) shown in Figs. \ref{fig15}(a)-(b), intense regions of opposite-signed vorticity form elongated structures around the lower and downstream portions of the cylinder and subsequently feed into the wake. This spatial organization is consistent with the strong rotational transport of vorticity at \(\Omega=5\). The moving wall continuously generates vorticity, while the imposed rotation advects the near-wall vorticity around the cylinder before it is released into the downstream flow. 

The higher modes progressively resolve the interaction between these near-body structures and the downstream wake. In particular, \(\phi_3\) and \(\phi_4\) shown in Fig. \ref{fig15}(c)-(d) retain the strong annular vorticity distribution around the cylinder while exhibiting more localized structures farther downstream. Modes \(\phi_5\) and \(\phi_6\) shown in Fig. \ref{fig15}(e)-(f) contain still more localized vortical concentrations in the near wake. The sequence consequently suggests a hierarchy in which the leading modes describe the large-scale rotation-induced vorticity organization, whereas higher modes account for the spatial modulation and breakdown of these structures as they are transported downstream.

The present structures are consistent with the enstrophy-based decomposition of supercritical cylinder wakes \cite{sengupta2015enstrophy}. \citet{sengupta2015enstrophy} showed that an enstrophy-based POD can resolve a hierarchy of instability-related spatial structures rather than reducing the dynamics to a single classical shedding mode. The present decomposition is likewise based on $\omega'$ and enstrophy, although the underlying flow is fundamentally modified by strong cylinder rotation. POD analysis of rotary-oscillating cylinders \cite{ping2020wake} has demonstrated that fewer modes suffice to reconstruct the near-wake two-layer shedding accurately such as in Fig. \ref{fig13} for $\Omega = 2.5$, while more modes must be retained to ensure an accurate approximation of the multi-scale temporal Magnus-Robins instability \cite{sundaram2021multiscale} reported for $\Omega = 5$.

\begin{figure}[!ht]
\centering
\includegraphics[width=.85\textwidth]{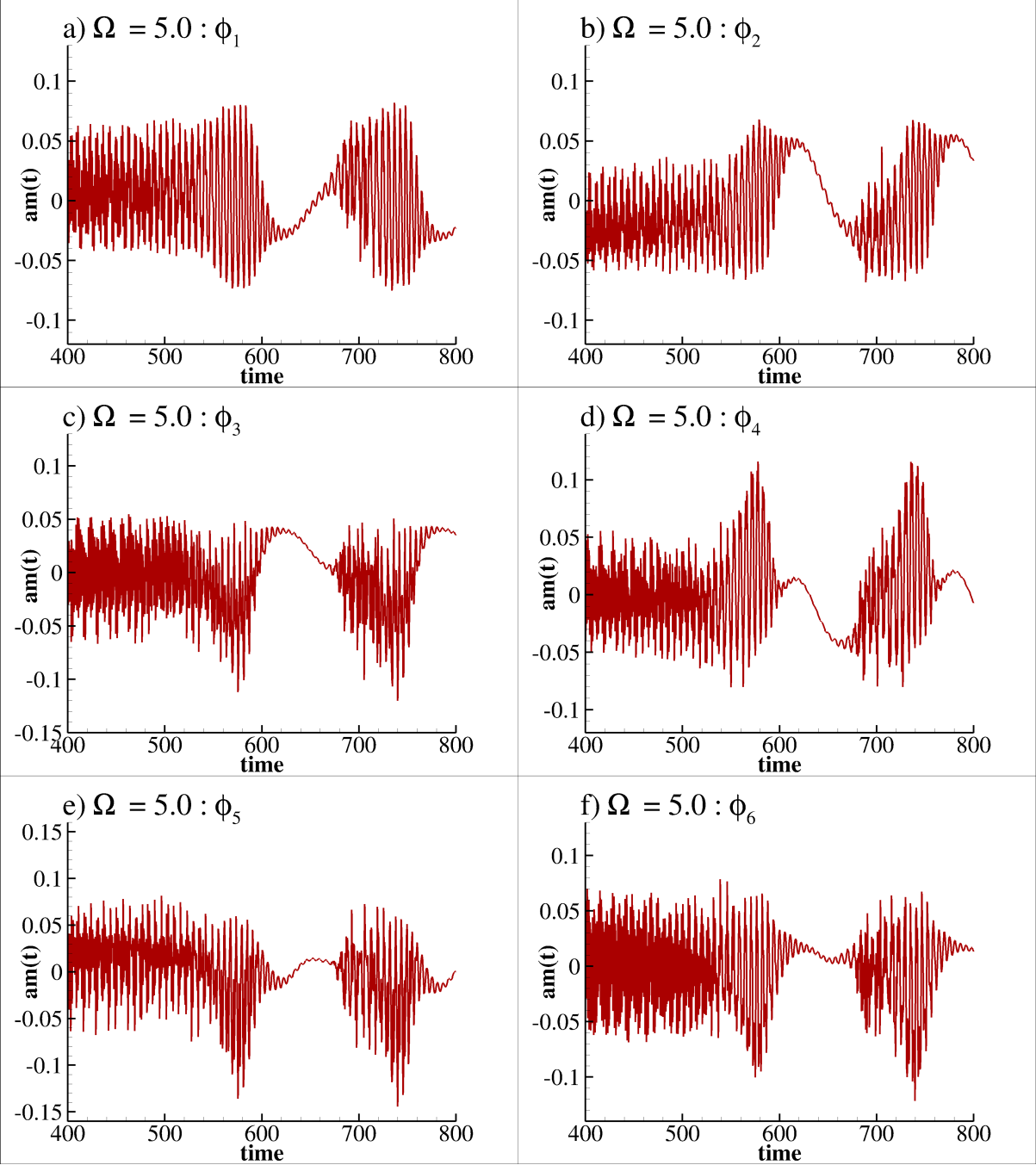}
\caption{Time-dependent amplitude functions of first six POD modes in frames (a)-(f), obtained for $\Omega = 5.0$ case, over time interval, $400 \le t \le 800$ in Fig. \ref{fig9}(c).}
\label{fig16}
\end{figure}

Figure \ref{fig16} presents the temporal coefficients of the first six POD modes at \(\Omega=5\). The modal amplitudes are generally within \(\pm 0.1\), with the largest excursions occurring for modes \(\phi_4\)–\(\phi_6\) shown in Fig. \ref{fig16}(d)-(f), reaching approximately \(\pm 0.1\)–\(0.13\). Unlike the more regular oscillations observed at \(\Omega=2.5\), all six coefficients exhibit strongly amplitude-modulated and intermittent oscillations, with periods of reduced activity separated by energetic bursts.

The temporal behaviour also suggests pairing among the modes, particularly \((\phi_1,\phi_2)\) and \((\phi_3,\phi_4)\), which exhibit comparable envelopes and oscillatory behaviour. Such complementary POD modes or \lq regular modes' are typically characteristic of travelling or oscillatory coherent structures \cite{joshi2025comparing}. Regular modes represent different phases of the same evolving structure, while their time coefficients describe its temporal phase progression. Similar paired POD modes have been reported for cylinder wakes and rotating/rotationally forced cylinder flows \cite{ping2020wake}. The similar temporal envelopes of \((\phi_1,\phi_2)\) and \((\phi_3,\phi_4)\) here suggest possible modal pairing. However, a definitive identification would require examination of their phase relationship or cross-correlation. For the \(\Omega=5\) case, the intermittent modulation of these modal coefficients indicates that the coherent structures themselves are not sustained periodically, but emerge and decay in bursts, consistent with the intermittent high-rotation instability observed in Fig. \ref{fig9}.

\begin{figure}[!ht]
\centering
\includegraphics[width=.85\textwidth]{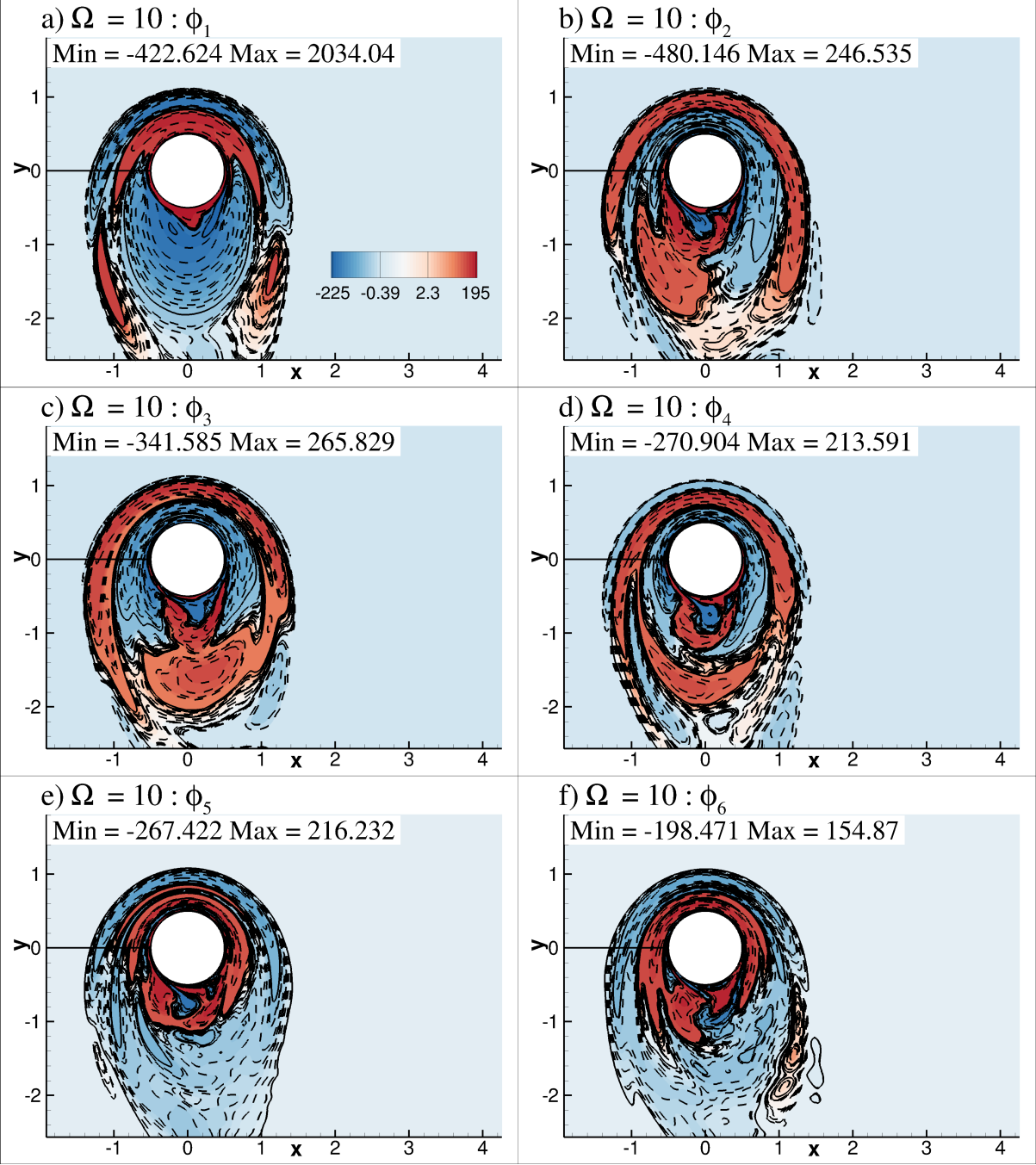}
\caption{Contour plots of first six spatial/eigenfunction modes ($\phi_1$ - $\phi_6$) from POD applied to $\Omega = 10.0$ case are shown in frames (a)-(f). The POD is evaluated over $400 \le t \le 800$ of Fig. \ref{fig9}(d).}
\label{fig17}
\end{figure}

Figure \ref{fig17} presents the first six eigenmodes of $\omega'$ for \(\Omega=10\). Here, the dominant vorticity is concentrated around the cylinder and form strongly wrapped, azimuthally organized structures \cite{suman2022novel} that extend into the downstream region. The six modes progressively separate the energetic near-cylinder rotational field from smaller, more localized wake structures. 

Mode $\phi_1$ shown in Fig. \ref{fig17}(a) is the dominant coherent structure and is concentrated in a closed, annular region surrounding the cylinder. The intense positive-vorticity region wraps around the upper and lateral portions of the cylinder, while opposite-sign vorticity appears mainly on the lower side and in the near wake. Thus, \(\phi_1\) represents the global rotation-dominated vorticity organization generated by the moving wall, rather than a conventional shed vortex. The second mode \(\phi_2\) shown in Fig. \ref{fig17}(b) has a similar spatial extent to $\phi_1$ but with a substantially different distribution of vorticity around the cylinder and in the lower wake. Similarity to \(\phi_1\) suggests that the two modes describe different phases/configurations of the same dominant rotation-induced coherent structure. For mode $\phi_3$ in Fig. \ref{fig17}(c), a pronounced coherent structure develops below the cylinder and extends into the near wake. Compared with \(\phi_1\) and \(\phi_2\), the mode places greater emphasis on the downstream redistribution of rotational vorticity, indicating that the POD hierarchy is beginning to separate near-wall rotational forcing from its wake response. Mode \(\phi_4\) in Fig. \ref{fig17}(d) retains the annular structure surrounding the cylinder but contains a more compact downstream region. It represents a more localized component of the near-body/wake interaction, than the global rotational field captured by the leading modes. For \(\phi_5\) shown in Fig. \ref{fig17}(e), vorticity becomes increasingly concentrated around the upper and lateral portions of the cylinder, with a distinct structure extending below it. This suggests a higher-order modulation of the rotation-induced shear layers and their subsequent transport into the wake. The sixth mode in Fig. \ref{fig17}(f) is weaker and more localized, with a compact structure near the lower-right downstream region. It is associated with smaller-scale or intermittent wake activity superposed on the dominant rotation-controlled flow.

\begin{figure}[!ht]
\centering
\includegraphics[width=.9\textwidth]{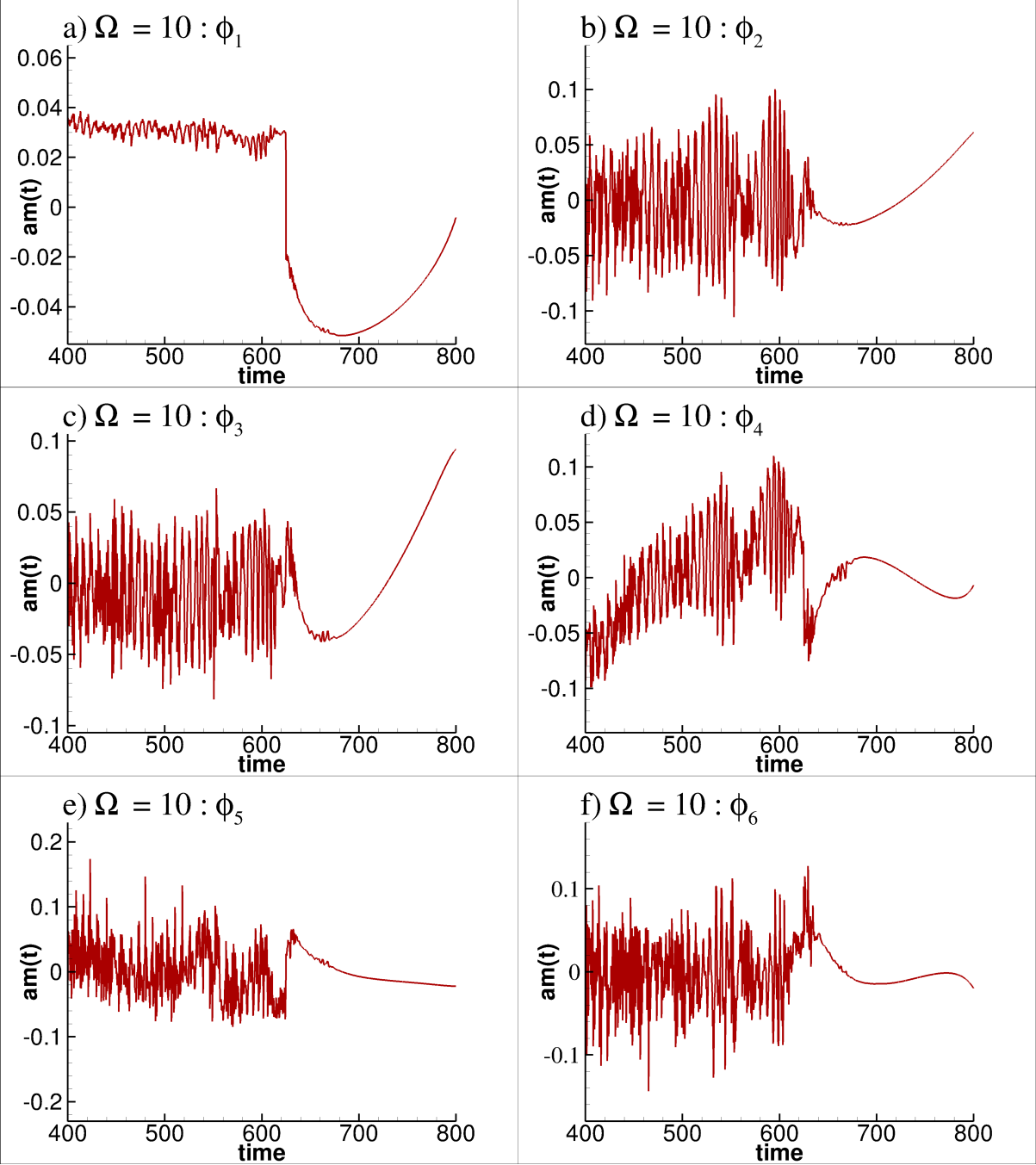}
\caption{Time-dependent amplitude functions of first six POD modes in frames (a)-(f), obtained for $\Omega = 10.0$ case, over time interval, $400 \le t \le 800$ in Fig. \ref{fig9}(d).}
\label{fig18}
\end{figure}

Figure \ref{fig18} presents the temporal evolution of the first six modal coefficients at \(\Omega=10\). During \(400\lesssim t\lesssim620\), all six $\phi$s exhibit pronounced high-frequency fluctuations, with amplitudes reaching approximately \(\pm0.03\) for \(a_1\), \(\pm0.1\) for \(a_2\)–\(a_4\), and up to \(\pm0.15\)–\(0.18\) for \(a_5\) and \(a_6\). The relatively large amplitudes of the higher modes indicate that the intense vortical activity during this interval involves a broad range of spatial scales rather than being dominated solely by the leading POD structures. A marked change occurs at approximately \(t=620\), after which the rapid fluctuations in all modal coefficients largely disappear. This behaviour directly corresponds to the inactive interval observed in the vorticity time history of Fig. 9(d). Unlike the classical periodic cylinder wake, no unambiguous \(90^\circ\) phase relationship is evident between the modal coefficients \cite{badr1990unsteady}. Although some modes exhibit similar amplitude envelopes during the active interval, this alone is insufficient to classify them as regular POD modes. The absence of a persistent phase-locked pair, together with the simultaneous burst and cessation of activity across multiple modes, is consistent with multiscale and intermittent characteristic of high-rotation unsteadiness \cite{sundaram2021multiscale, sengupta2025bifurcation}. 

\section{Summary and conclusions}
The present study examined the influence of rotation on the flow physics of a circular cylinder in two-dimensional compressible flow at \(Re_\infty=3800\) and \(M_\infty=0.1\). By systematically varying the rotation rate over \(0.1\leq\Omega\leq10\), the study demonstrates that rotation does not simply weaken the conventional cylinder wake monotonically. Instead, it reorganizes the flow through a sequence of qualitatively different dynamical states, involving the suppression of classical vortex shedding and the subsequent emergence of a distinct high-rotation unsteadiness.

At low rotation, the flow is governed by the familiar interaction between the two separated shear layers and the formation of an alternating vortex street. Increasing rotation progressively breaks this symmetry by imposing a strong velocity difference between the two sides of the cylinder. The resulting imbalance in vorticity generation and transport weakens the conventional shedding process and eventually leads to a nearly steady wake. The transition is particularly narrow around \(2.2\lesssim\Omega\lesssim2.8\), where the temporal response changes from sustained oscillations to intermittent activity before becoming largely suppressed. This suggests that the intermediate regime is not merely a gradual reduction in shedding amplitude, but a reorganization of the dominant wake dynamics.

Further increase in rotation produces a second transition. Unsteadiness reappears at higher rotation rates, but its temporal character is markedly different from that of the low-rotation wake. The intermittent bursts in lift and local vorticity, together with the broadband spectral response, indicate that the re-emergent instability is not a continuation of the classical Kármán vortex shedding modes. Instead, the flow is dominated by the interaction between intense rotation-induced near-body vorticity and its asymmetric downstream transport. The increasing onset time with rotation further suggests that stronger rotational forcing progressively delays the development of this instability, even though sufficiently high rotation eventually supports its re-emergence.

The surface distributions provide an additional physical perspective on this transition. Increasing rotation substantially modifies both the pressure and streamwise wall-shear distributions, with the increasingly asymmetric pressure field providing the principal physical mechanism for the increasing Magnus-Robins lift. The non-monotonic behaviour of drag, particularly its enhanced variability at high rotation, indicates that the mean aerodynamic response cannot be inferred solely from the increasing rotational velocity. Instead, it reflects the competing effects of rotation-induced acceleration, wake modification and intermittent vortex activity. An important consequence is that appreciable local compressibility develops despite the low $M_{\infty}$. Rotational acceleration produces local Mach numbers approaching 0.5. Thus, the relevant flow physics are governed by the local flow speed generated by rotation rather than by \(M_\infty\) alone. 

The vorticity-based POD analysis provides a complementary description of this evolution. The dominant modes change from extended wake structures at low rotation to strongly wrapped near-cylinder structures at high rotation. The reduction in effective dimensionality near the suppressed $\Omega$ regime indicates that the remaining vortical field becomes spatially organized despite the persistence of weak local fluctuations. Conversely, the increasing modal complexity at high rotation reflects the coexistence of energetic near-body structures and localized downstream structures. The temporal coefficients further show that these structures are intermittent rather than demonstrating a persistent phase-locked oscillation. Hence, the POD results support the interpretation that the high-rotation regime represents a distinct multiscale dynamical state, rather than a recovery of the low-rotation shedding mechanism.

Overall, the results show that rotation acts not only as a means of modifying the strength and direction of the cylinder wake, but as a control parameter that reorganizes the underlying hierarchy of coherent structures and temporal dynamics. The resulting sequence of shedding, suppression and high-rotation re-emergence provides a unified framework for interpreting the aerodynamic and vortical response of the rotating cylinder. Further investigation using global stability analysis and three-dimensional simulations would be valuable for determining the precise instability mechanisms responsible for the high-rotation regime and assessing the persistence of the observed transitions beyond the present two-dimensional framework.

\begin{acknowledgments}
The authors would like to acknowledge the use of high-performance computing facilities CHANDRASEKHAR and ARYABHATA at Indian Institute of Technology Dhanbad for all the computations reported here. 
\end{acknowledgments}

\section*{Declaration of Interests}
The authors report no conflict of interest.

\section*{Data Availability}
The datasets produced from the high-fidelity simulations are available from the corresponding author upon reasonable request. 


\bibliography{rot_cyl}


\end{document}